# From biodiversity modelling to conservation action: a spatial indicator for prioritising tropical forest protection, restoration, and management

Maïri Souza Oliveira[a,b]*, Maxime Lenormand[a*], Sandra Luque[a]

[a]INRAE, National Research Institute on Agriculture, Food & the Environment, TETIS research unit, Maison de la télédétection, 34090 Montpellier, France

[b]CIRAD, Centre for International Cooperation in Agricultural Research for Development, Forests and Societies research unit, Montpellier 34398, France

*Corresponding author: mairisouzaoliveira@gmail.com and maxime.lenormand@inrae.fr

## Graphical abstract

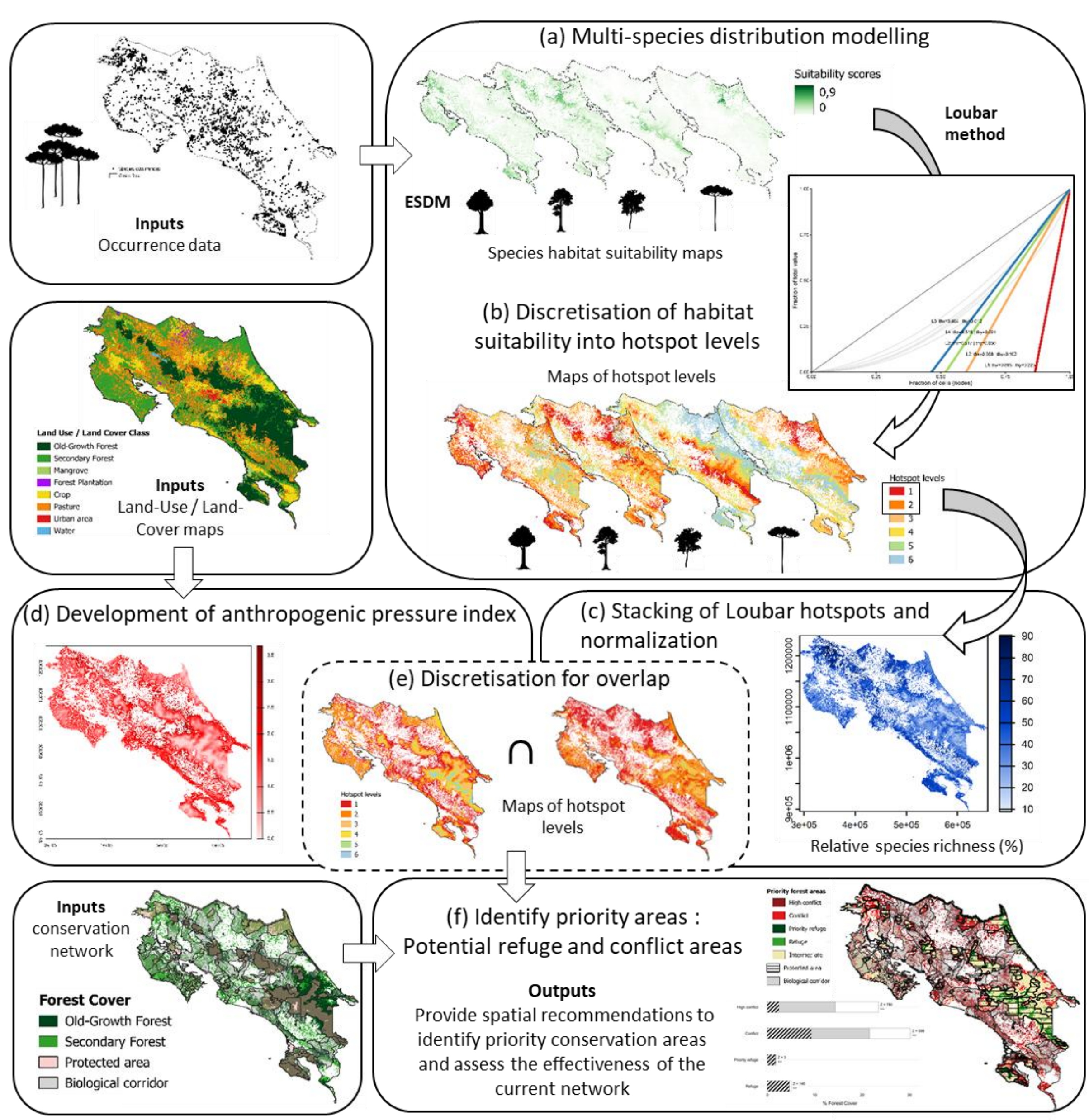

## Abstract

Tropical forests harbour exceptional biodiversity but are increasingly threatened by anthropogenic pressures, making their conservation central to achieving the Convention on Biological Diversity targets. However, national conservation planning is often constrained by heterogeneous monitoring frameworks and limited integration of biodiversity and anthropogenic pressures. We develop a synthetic spatial indicator that combines open-access biodiversity and land-use data to identify refuge and conflict areas between ecological potential and anthropogenic pressure. Using Costa Rica as a national-scale demonstrator, we assess its capacity to support conservation prioritisation and evaluate the current conservation network. Ecological potential was estimated using multi-species distribution models for 254 dominant canopy tree species, serving as an operational proxy for tropical forest biodiversity and ecological structure. We quantified anthropogenic pressure from land-use patterns and combined it with ecological potential to identify refuge and conflict areas. We then tested the spatial significance of these areas and evaluated their representation within the national conservation network. The indicator reveals a clear spatial organisation of biodiversity–pressure interactions: refuge areas coincide with large, continuous forest cores, whereas conflict areas are concentrated in fragmented landscapes and may extend into protected areas. Flexible and transferable, the framework distinguishes areas requiring strict protection from those where restoration, sustainable management, or pressure mitigation should be prioritised. Rather than prescribing conservation actions, the refuge–conflict indicator provides a practical decision-support tool to assess conservation networks, strengthen ecological connectivity, and guide national biodiversity planning.

## 1. Introduction

Tropical forests harbor a major share of terrestrial biodiversity, while being subject to increasing pressures that are driving an alarming erosion of this biodiversity (Barlow et al., 2018; Gibson et al., 2011). Biodiversity loss is primarily due to human activities leading to land conversion, landscape fragmentation and degradation, and climate change (Caro et al., 2022; Curtis et al., 2018; FAO, 2025; Pendrill et al., 2022). Recent syntheses show that, beyond deforestation, the integrity and connectivity of remaining large contiguous forest areas are rapidly declining, with direct implications for species persistence and ecosystem functioning (Grantham et al., 2020). These forests are central to conservation strategies aimed at achieving the objective of the Convention on Biological Diversity (CBD – COP15), which seeks to ensure the effective protection and restoration of at least 30% of ecosystems by 2030 (CBD, 2022; Mrema et al., 2020). In this context, spatial planning must move beyond identifying priority areas and instead establish actionable priorities to guide national-scale decisions between protection (preventing loss) and restoration (recovering ecological potential) (Chazdon et al., 2024; Marshall et al., 2022; Possingham et al., 2015). The restoration ecological success depends strongly on spatial targeting, and explicit prioritisation can significantly alter the effectiveness of investments (Holl et al., 2022; Strassburg et al., 2020). However, operationalising robust prioritisation frameworks remains difficult at national scale in tropical contexts, for many reasons.

On the one hand, approaches centred on rarity or threat status (e.g. the IUCN Red List) remain constrained by taxonomic biases and data deficiencies, particularly marked for plants and many tropical groups. Recent studies document these limitations and their consequences for extinction risk assessment and for the usability of assessments in decision-making (Borgelt et al., 2022; Thomas et al., 2024). Moreover, even when sufficient occurrence data are available, focusing solely on endemic or threatened species is insufficient. To preserve ecosystem functions, particularly structural ones, and to avoid biasing conservation strategies, it is necessary to incorporate all species, including those that are less represented or not threatened (Akçakaya et al., 2020; Rey et al., 2024; Zhang et al., 2022). Finally, frameworks such as Key Biodiversity Areas (KBA – IUCN, 2016), based on criteria encompassing the presence of threatened species, restricted-range taxa or ecosystems, intact ecosystems, essential ecological processes or demonstrated irreplaceability (Plumptre et al., 2025), although structuring, often require homogeneous datasets and implementation capacity that are not always available at national level. Moreover, the question of ecological condition of sites (forest integrity) becomes central to making these frameworks truly operational (Crowe et al., 2023).

Many prioritisation approaches for conservation rely primarily on ecological value or importance, whereas the explicit and joint integration of a spatial assessment of pressures (threats, disturbances, vulnerabilities) remains uneven and is not undertaken systematically (Grantham et al., 2020; Pla et al., 2024; Ward et al., 2020). Yet forest integrity approaches show that a significant proportion of remaining forests is already highly altered, including within protected areas. The simple presence of forest (quantitative state) does not guarantee high integrity or realistic conservation potential (qualitative state) (Edgar, 2025; Watson et al., 2018). Without coupling ecological value and anthropogenic pressure, prioritisation maps risk being of limited interpretability in terms of measures (protect vs restore vs actively manage).

Finally, a further challenge concerns the representation of biodiversity at large scale. The use of proxies is often indispensable, but their performance and limitations must be made explicit (Barlow et al., 2007). Recent analyses of Neotropical conservation prioritisation reveal spatial mismatches between priorities based on composition, structure, and function, highlighting the value of integrative proxies and the need to test their effectiveness rather than assume it (Burbano-Girón et al., 2022). In tropical forests, where exhaustive coverage of diversity is unrealistic, assemblages of dominant canopy species offer strong potential as both an operational and ecologically grounded proxy. These abundant species structure forest architecture and, at broader scales, ecosystems (Souza Oliveira et al., 2025a), modulate microclimate, and integrate broad biogeographical and edaphic gradients (Lohbeck et al., 2016; Rüger et al., 2023; ter Steege et al., 2013). Their composition synthesises functional strategies and controls major processes (carbon, water, energy, light, litter), thereby conditioning a substantial share of the environmental filters that structure the composition and realised niches of lower strata (Aguirre-Gutiérrez et al., 2025b; Balvanera et al., 2021). Changes in dominant assemblages often provide early signals of degradation or transformation, particularly in anthropised landscapes (Barlow et al., 2018). Thus, the spatial distribution of dominant species provides a coherent, continuous and updateable signal to guide prioritisation (protection vs restoration), in complement to inventories, structural indicators and field validation (Cazzolla Gatti et al., 2022; Lindenmayer et al., 2015).

Against this background, we propose an operational national prioritisation approach that translates potential multi-species distributions into a spatial indicator designed to support biodiversity conservation planning and implementation of the Kunming–Montreal Global Biodiversity Framework (GBF). In particular, the indicator is intended to inform Targets 1 (integrated spatial planning), 2 (ecosystem restoration), and 3 (effective conservation of ecologically important areas), by providing a transparent and reproducible basis for identifying areas of high conservation value and evaluating the adequacy of existing conservation networks (Unit, 2024).

The objective is to assess the extent to which a synthetic indicator, built from open-access biodiversity and land-use data together with a structuring ecological proxy, can support national conservation prioritisation and gap analyses of protected-area networks. The originality of the approach lies in the explicit integration of (i) potential ecological value derived from multi-species distribution modelling of canopy-structuring species and (ii) a gradient of anthropogenic pressure inferred from land-use patterns. By combining these complementary dimensions, the indicator distinguishes refuge areas (high ecological potential, low anthropogenic pressure), representing priorities for long-term protection, from conflict areas (high ecological potential, high anthropogenic pressure), where restoration, sustainable management, or mitigation actions are most urgently needed. Finally, we evaluate the complementarity of these priorities with the existing network of protected areas and ecological corridors to assess the effectiveness of the national conservation network and identify opportunities for improving its contribution to GBF implementation.

We implement this framework proposing workflow to be tested in Costa Rica (Figure 1a) as a national-level demonstrator to illustrate how a reproducible spatial indicator can support biodiversity reporting and conservation planning under the GBF. Designed from open-access biodiversity and land-use data, the approach is intended to be transferable to other tropical countries seeking transparent methods for identifying conservation priorities, evaluating the representativeness of protected-area networks, and supporting reporting on GBF Targets 1–3. Costa Rica provides a particularly relevant testing context because its strong climatic and altitudinal gradients generate a complex mosaic of tropical forest ecosystems, ranging from dry forests to montane wet forests (Souza Oliveira et al., 2025a). Its forest landscapes also encompass contrasting conservation situations. Old-growth forests are mainly concentrated in protected mountainous areas and secondary forests account for approximately 36% of national forest cover (Figure 1b). They are often fragmented and embedded within agricultural, pastoral, and monoculture-dominated matrices (Figure 1c - Montero et al., 2021). This configuration reflects a dynamic land-use trajectory, marked by intensive deforestation during the 1970s–1980s, subsequent forest recovery from the 1990s onward, and the implementation in 1996 of a regulatory framework prohibiting the conversion of public and private forest lands, although early successional stages remain imperfectly covered by this protection regime (Arroyo-Mora et al., 2014; Shaver et al., 2015). These ecological and socio-environmental contrasts make Costa Rica an ideal national-scale demonstrator for assessing whether the proposed indicator can consistently identify areas requiring protection (refuge areas), restoration, or improved management (conflict areas), while simultaneously evaluating the contribution of existing protected areas and ecological corridors to national conservation objectives and GBF implementation.

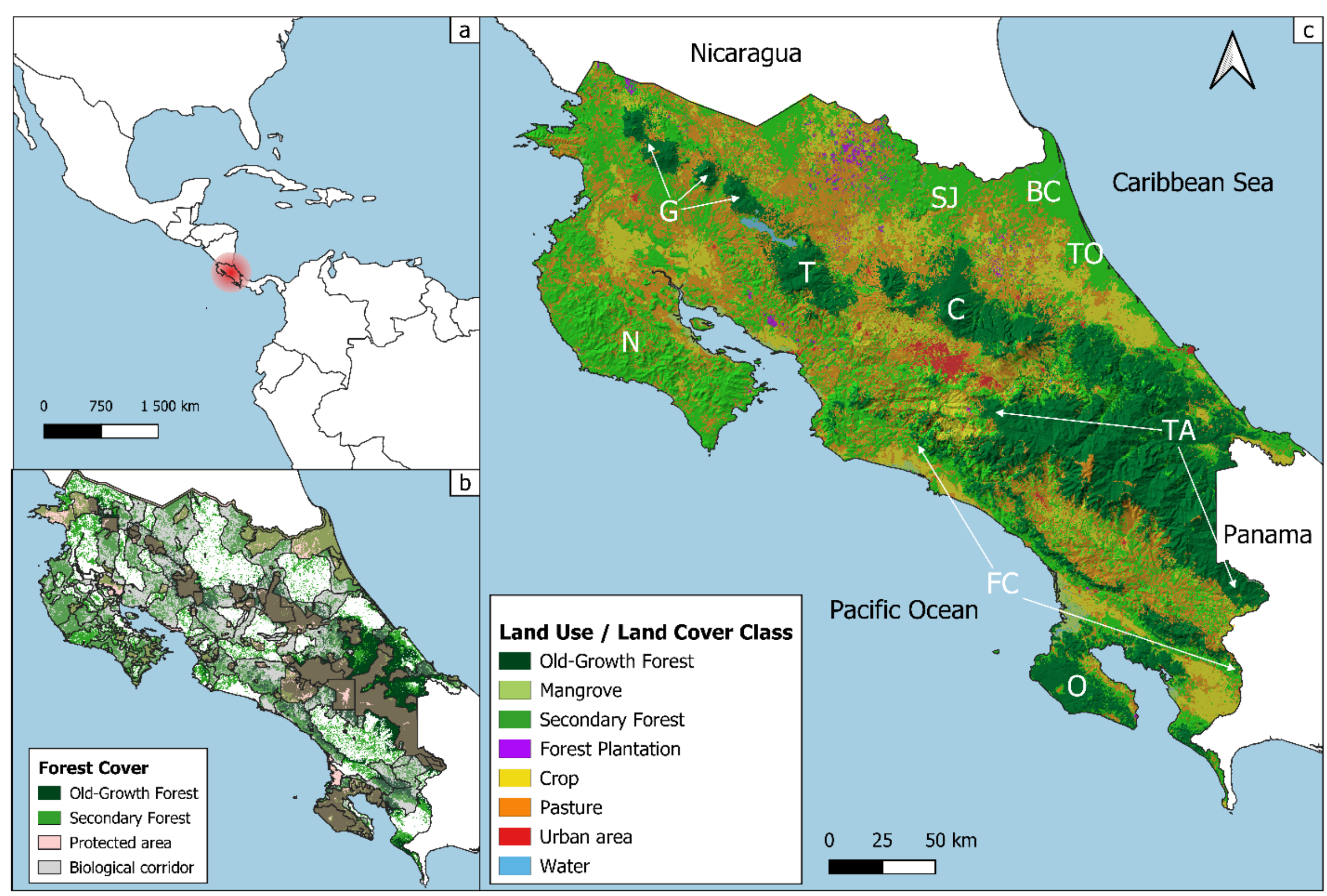

**Figure 1.** Spatial data context used to implement the national prioritisation framework in Costa Rica: (a) country's location in America, (b) The national forest cover in 2021 with the boundaries of protected areas, (c) reclassified SINAC land-use/land-cover map for 2017–2021 with the location of interest areas : Barra del Colorado (BC), Central mountain range (C), Fila Costeña Range (FC), Guanacaste mountain range (G), Nicoya peninsula (N), Osa peninsula (O), Biological corridor San Juan - La Selva (SJ), Tilarán mountain range (T), Talamanca mountain range (TA), Toruguero (TO).

## 2. Material and Methods

We therefore develop a reproducible spatial indicator designed to support national conservation planning and GBF reporting by combining open-access biodiversity and land-use land-cover data within a prioritisation framework (Figure 2). The indicator is implemented in Costa Rica as a national-scale demonstrator to assess its ability to identify refuge and conflict areas, evaluate their complementarity with protected areas and ecological corridors, and illustrate its transferability to other tropical countries.

### 2.1. Data

We compiled data for 367 key canopy-structuring species from the main forest ecosystems identified at the national scale in Souza Oliveira et al. (2025a) using the GBIF database (GBIF.org, 2025a-j). We extended the temporal coverage of the data from 2004 to 2025 to ensure a sufficient number of occurrences for modelling and to align with the period of the forest inventories used to delineate the main ecosystems. This approach assumes relative stability of forest cover since the 2000s, based on the Forest Law No. 7575 of 1996 (MINAE, 1996). In total, we used 18,075 occurrences, ranging from 10 to 193 per species. Details on species and occurrence numbers are provided in Figure S1 and Table S1.

To assess the potential distribution of the studied species, i.e. their abiotic ecological niche, we selected variables representing vegetation dynamics, topography, soil, and climatic conditions, as previously used in ecosystem modelling by Souza Oliveira et al. (2025a), to which the species studied contribute significantly:

- Vegetation: The Normalized Difference Water Index (NDWI) during the wet season (November to January) was calculated as the median over the 2018 multitemporal optical image series from the Copernicus Sentinel-2 L2A satellite, at 10 m and 20 m spatial resolution (CEOS, 2024), available as open access ("costarica-sentinel-2-l2-spectral-indices," n.d.).
- Topography: Elevation (m) and slope (°) were derived from the NASADEM Merged DEM Global 1 Arc-Second V001, with a spatial resolution of 30 m (NASA JPL, 2020).
- Climate: Mean annual precipitation (anPR, mm/year) and the coefficient of variation of precipitation seasonality (PRSea, %), at 1 km resolution, were obtained from CHELSA v1.0 (Karger et al., 2017).
- Soil: topsoil Cation Exchange Capacity (CEC30, mmolc/kg) and pH (pH30), at 250 m resolution, were sourced from SoilGrids v0.5.5 (Hengl et al., 2017).

To develop a spatial indicator of local anthropogenic pressure on forest cover, we calculated metrics at the class level from the Costa Rica LULC map at 10 m spatial resolution produced by SINAC (SINAC, 2021). This map was generated from a classification of Sentinel-2 and Sentinel-1 mosaics and supplemented with the 2017 LULC map for missing pixels. It was reclassified and aggregated to 250 m to ensure consistency with the resolution of habitat suitability maps (Figure 1c). The spatial layers of protected areas and biological corridors used (Figure 1b), as well as the LULC data are available freely via SNIT ("SNIT," n.d.).

## 2.3. Ensemble Species Distribution Models from environmental data

To identify priority areas for the studied species, we used a multi-species workflow based on species-level ensemble species distribution models (ESDM) fitted to cleaned occurrence records compiled from GBIF (Figure 2a) (GBIF.org, 2025a-j), which aggregates heterogeneous biodiversity observations including citizen-science contributions (Araújo and New, 2007; Zizka et al., 2020; Zwiener and Alves, 2023).

Environmental variables were resampled to a common spatial resolution of 250 m, with upscaled layers (NDWIw, DEM, and slope) aggregated using mean, standard deviation, and maximum values. To limit overfitting, a variable selection procedure was applied in which one variable from each group of collinear variables was selected based on Pearson correlation ($r \geq 0.7$). The same variables were selected for all species. The relative contribution of each variable, calculated independently of the modelling algorithms, corresponds to the Pearson coefficient between the initial model predictions and those obtained via a jackknife omission test of the evaluated variable (Hawkins et al., 2017).

To reduce sampling bias associated with multiple occurrences within the same pixel, presence points were spatially rarefied to match the resolution of the environmental and species data (250 m) and to avoid clustering, retaining only one randomly chosen occurrence per pixel (Souza Oliveira et al., 2025b). Following this filtering, 6,823 occurrences were retained for 254 species. The resulting species list and the number of occurrences at this spatial resolution are presented in Table S1.

For each species, 1,000 pseudo-absences were randomly generated within the background using disk stratification to avoid spatial clustering around presence points. Presences and pseudo-absences were weighted equally in the model. This approach ensures a balanced representation of presence and absence data for the algorithms included in the ESDM.

Models were run using seven algorithms available in the SSDM R package (Schmitt et al., 2017) to improve robustness, accuracy, and stability of SDMs, accounting for uncertainty related to algorithm choice (Kaky et al., 2020; Montoya-Jiménez et al., 2022; Zurell et al., 2020). The algorithms used were: Generalised Additive Model (GAM), Generalised Linear Model (GLM), Multivariate Adaptive Regression Splines (MARS), Artificial Neural Networks (ANN), Classification Tree Analysis (CTA), Random Forest (RF), and Maximum Entropy (MAXENT). Each model was calibrated ten times and validated using holdout cross-validation, with 70 % of the data used for training and 30 % for validation, over five iterations. Habitat suitability maps for each species were produced from an ensemble model, calculated by weighting the suitability scores of individual models with an Area Under the ROC Curve (AUC) ≥ 0.7 (Hanley and McNeil, 1982).

Detecting false positives and false negatives is essential for biodiversity conservation, as overestimating or underestimating a species' potential range can lead to inappropriate management decisions (Guisan et al., 2013). We retained 119 species with a low number of occurrences (10 to 20) to include localized taxa that may be important for maximizing hotspot detection within our overall workflow (Konowalik and Nosol, 2021; Sampaio and Cavalcante, 2023; van Proosdij et al., 2016). In this context, no single metric is reliable, particularly with small samples; therefore, it is recommended to combine indicators of discrimination, calibration, bias, and accuracy (Konowalik and Nosol, 2021). To assess model quality, we selected a complementary set of three metrics suitable for adequacy scores and underrepresented species: (i) the AUC, which measures a model's ability to discriminate presence sites from absence sites; (ii) the Continuous Boyce Index (CBI), a calibration metric, which evaluates the accuracy of occurrence predictions relative to a random distribution of occurrences along the prediction gradient; and (iii) Bias, which measures the systematic deviation between predicted and observed values, calculated as the mean difference (Konowalik and Nosol, 2021; Zurell et al., 2020) (Konowalik and Nosol, 2021; Zurell et al., 2020). The ecological consistency of habitat suitability maps for species whose metrics may indicate poor model performance was evaluated through expert assessment (Zurell et al., 2020).

## 2.4. Development of a local anthropogenic pressure indicator derived from LULC data

To characterise local anthropogenic pressure around forest pixels, we developed a landscape-based pressure indicator derived from LULC data by combining class-level landscape metrics describing the composition and configuration of anthropogenic land-use classes (Figure 2d).

Anthropogenic pressures on forests, at both landscape and local scales, determine their ecological state and are primarily influenced by the composition and configuration of the surrounding landscape, which affect connectivity, edge effects, and biodiversity (Arroyo-Rodríguez et al., 2020; Fahrig, 2017, 2013; Rosenfield et al., 2023). Landscape metrics describing anthropogenic land-use classes thus provide robust spatial indicators to characterise these pressures and to identify conflict zones and biodiversity refuges. Fifteen class-level landscape metrics were calculated to quantify the importance and impact of each anthropogenic class (forest plantations, croplands, pastures, urban areas) on the landscape surrounding each forest pixel. The full set of metrics was selected based on their definitions in FRAGSTATS (McGarigal et al., 2022) and are presented in Table S2. We selected the focal window size for moving-window landscape metrics using a cross-scale sensitivity analysis based on three complementary indicators: (i) a spatial roughness index, (ii) global variance, and (iii) inter-scale stability quantified by correlation with the smallest window. Candidate windows, ranging from 1 km to 10 km, were chosen with reference to spatial scales documented in the scientific literature at which anthropogenic pressures significantly affect forest patch floristic patterns, while remaining appropriate for the 250 m resolution of the data used (Grantham et al., 2020; Haddad et al., 2015). Full equations and implementation details are provided in Section S1.

Among the fifteen landscape metrics calculated, a collinearity analysis was conducted to select a set of non-collinear ($|r| < 0.8$) and complementary metrics, identified in the literature as relevant, in order to characterise the importance and influence of each class on landscape structure. The same selection was applied to all anthropogenic classes to produce comparable and subsequently combinable indices. The metrics were then standardised (z-score) and some transformed (reoriented) to ensure directional consistency across all indicators, so that increases in their values consistently corresponded to intensification of anthropogenic pressure. An independent PCA was conducted for each anthropogenic class using the standardised and directionally aligned metrics. When the first principal component explained more than 70% of the total variance, it was retained as the class-specific pressure score. Otherwise, a weighted combination of the leading components was calculated, with weights proportional to their explained variance, retaining enough components to reach 70% cumulative variance. The axes were reoriented so that increasing scores consistently reflected an intensification of anthropogenic pressure. The resulting scores were then normalised between 0 and 1 to produce a continuous index for each class, and subsequently summed to obtain the landscape-scale pressure index $P(x)$, bounded within [0,4] .

## 2.5. Discretisation into "hotspot" levels: construction of the Indices *b* and *p*

Continuous spatial layers generated throughout the workflow, namely habitat suitability scores, relative species richness, and anthropogenic pressure, were represented as raster maps. Because these variables differed in scale, range, and distribution, and often showed highly skewed spatial patterns, direct comparison or overlay was not straightforward. To preserve the information contained in the continuous gradients while defining spatially comparable classes, we reclassified these rasters into ordered hotspot levels using the Loubar method (Louail et al., 2014), applied iteratively (Bassolas et al., 2019; Préau et al., 2022) (Figure 2b, 2e).

Let $Z(x)$ denote the value of a continuous raster at pixel $x$. The Loubar method relies on the analysis of the Lorenz curve of Z and on the tangent to this curve at point (1,1), allowing the non-parametric identification of thresholds that isolate the values concentrating most of the signal. By iteratively applying this procedure, each continuous raster was converted into six ordered hotspot levels:

$H_z(x) \in \{1,2,3,4,5,6\}$ (1)

where $H_z(x) = 1$ corresponds to the strongest hotspot, i.e. the highest values of $Z$, and $H_z(x) = 6$ corresponds to the lowest values.

### 2.5.1. Estimation of continuous relative species richness

For each species $s \in \{1, \dots, S\}$, the ESDM produced a continuous habitat suitability map. Rather than binarising outputs, we identified areas of high suitability for each species using Loubar reclassification applied to habitat suitability scores (Figure 2b). This provides a species-specific ordinal suitability map:

$H_s(x) \in \{1,2,3,4,5,6\}$ (2)

where $H_z(x) = 1$ corresponds to pixels with the highest suitability (niche core), levels 2–5 represent a decreasing gradient surrounding this core, level 6 groups pixels with very low suitability. This approach preserves species-specific suitability gradients while standardising their distributions into ordinal classes, enabling comparison and overlay of maps with very different spatial scales.

For each species, uncertainty associated with the probability of presence map was propagated using a Monte Carlo approach. Based on the mean and standard deviation provided by the ESDM, 1,000 random draws were generated for each pixel following a truncated normal distribution bounded between 0 and 1. These simulations allow assessment of variability in presence probabilities and quantification of confidence in the final hotspot level assignment using the Loubar method. The most frequent hotspot class across simulations was retained as the mode, and the proportion of concordant iterations provided a confidence index for each pixel.

This approach combines the central information from the ESDM with uncertainty propagation, ensuring that areas identified as hotspots best reflect the true probability of presence for each species (Araújo et al., 2019; Thuiller et al., 2019; Zurell et al., 2020). We then defined a binary high suitability indicator for each species:

$$I_s(x) = \{1,\ if\ H_s(x) \in \{1,2\}\ 0,\ else \quad (3)$$

Relative species richness was then estimated by stacking species-level high-suitability indicators:

$$B(x) = \frac{1}{S}\sum_{s=1}^{S} I_s(x) \times 100 \quad (4)$$

where $B(x)$ represents the percentage of modelled species exhibiting high suitability at pixel $x$. This step provides a continuous multi-species metric interpretable as relative potential species richness..

**2.5.2. Combining anthropogenic pressure and potential biodiversity hotspot levels**

To directly compare and overlay potential biodiversity and anthropogenic pressure on a common scale, we converted the two continuous variables $B(x)$ and $P(x)$ into ordinal hotspot levels using the Loubar method, and analysed their overlap at the pixel scale (Figure 2e).

Relative species richness and anthropogenic pressure are described respectively by continuous rasters $B(x)$ bounded within [0,100], and $P(x)$, bounded within [0,4]. Loubar discretisation was applied separately to each raster:

$$b(x) = H_B(x), b(x) \in \{1,2,3,4,5,6\} \quad (5)$$

$$p(x) = H_P(x), p(x) \in \{1,2,3,4,5,6\} \quad (6)$$

where $b(x) = 1$ indicates pixels concentrating the highest biodiversity values while$b(x) = 6$ corresponds to the lowest ; (ii) Similarly, $p(x) = 1$ to pixels experiencing the strongest anthropogenic pressure, whereas$p(x) = 6$ indicates the weakest pressure..

Each pixel $x$ is thus characterised by an ordinal pair;

$$\big(b(x), p(x)\big) \in \{1, \dots ,6\} \times \{1, \dots ,6\} \quad (7)$$

corresponding to36 possible combinations. This cross-classification defines a discrete space describing the co-variation between potential biodiversity and anthropogenic pressure.

## 2.6. A posteriori calibration of "conflict/refuge" categories and assessment of spatial significance

Refuge and conflict categories (and their intensities) were not defined a priori. They were determined a posteriori from the observed structure when the combinations $\big(b(x), p(x)\big)$ were projected into the continuous space $\big(B(x), P(x)\big)$. We examined the joint distribution of the 36 combinations and their relative positioning in the (B,P) plane in order to identify coherent groupings corresponding to conflict areas (high potential biodiversity and high pressure), refuge areas (high potential biodiversity and low pressure), and intermediate situations. This calibration ensures that the ordinal classes $b(x)$ and $p(x)$ correspond to interpretable groupings in continuous space $(B, P)$, and defines synthetic categories (refuge/conflict) adapted to the observed distributions within the study area (Figure 2f).

After calibration, the spatial significance of these zones was assessed using a shuttle-type permutation test (500 iterations), providing a Z-score for each class that measures deviation from randomness. The proportion of permutations showing an overlap equal to or greater than the observed constitutes a one-tailed p-value, indicating whether the observed spatial configuration is significantly clustered or dispersed.

All analyses were performed using R version 4.3.3 (R Core Team, 2024). Ensemble modelling was conducted with the SSDM package version 0.2.11 (Schmitt et al., 2017), and landscape metrics were calculated using the landscapemetrics package version 2.2.1 (Hesselbarth et al., 2019).

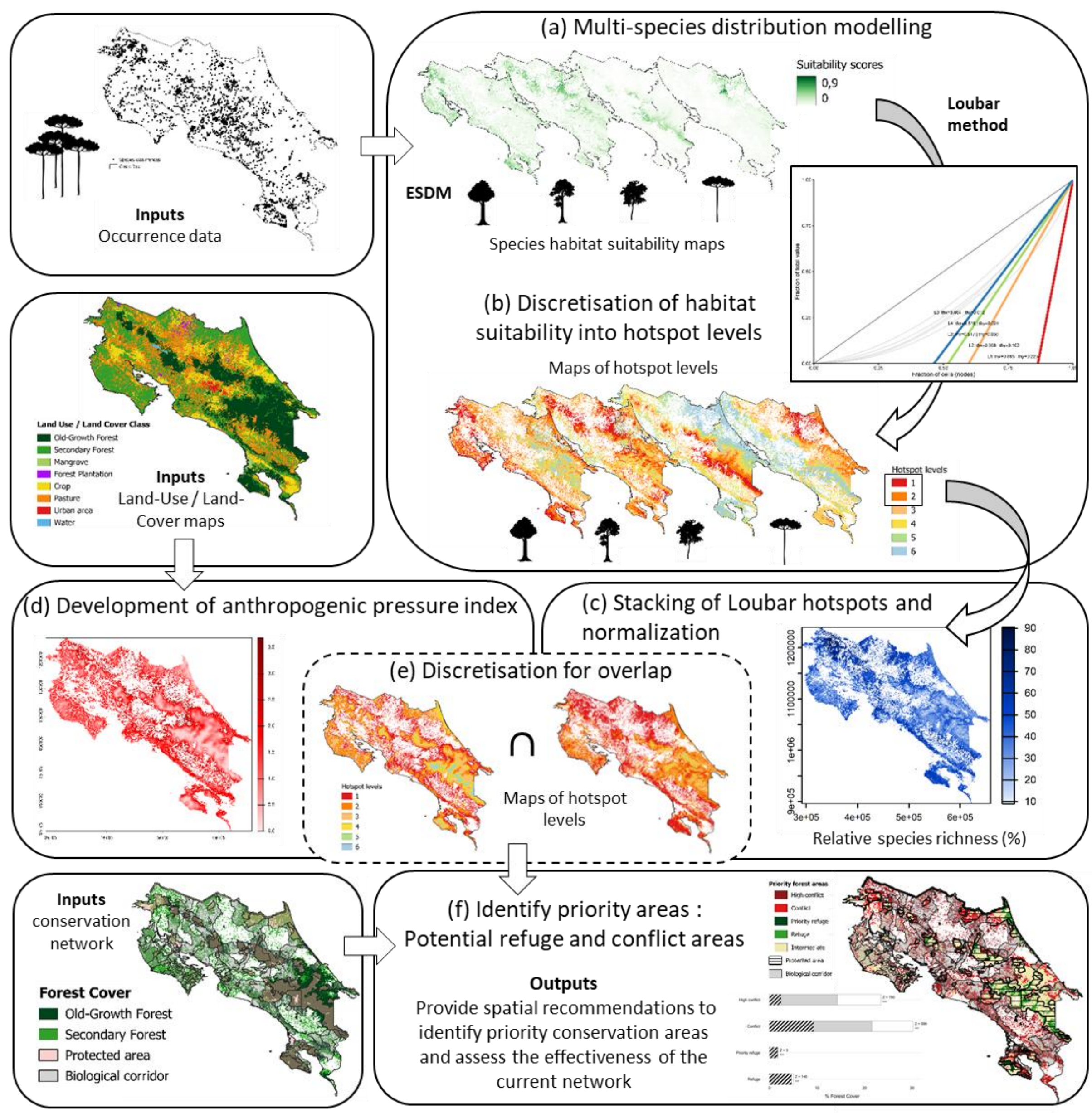

**Figure 2.** Schematic representation of the spatial indicator and spatial prioritisation workflow, illustrating the key methodological steps.

# 3. Results

## 3.1. Assessment of Ensemble Species Distribution Models from environmental data

All 254 species were successfully modelled, with final ESDMs selected using an AUC threshold of 0.7. Model performance was overall satisfactory, with high AUC values (0.70–0.95), low bias (< 0.18), and generally good calibration despite substantial interspecific variation in CBI (Figure S2). Five species with low calibration performance (CBI < 0.5) were further assessed through expert evaluation of habitat suitability maps to verify the ecological consistency of predictions. Figure 3 illustrates the range of ESDM outputs and the Loubar-based hotspot discretisation using four representative species with contrasting sample sizes and model performance.

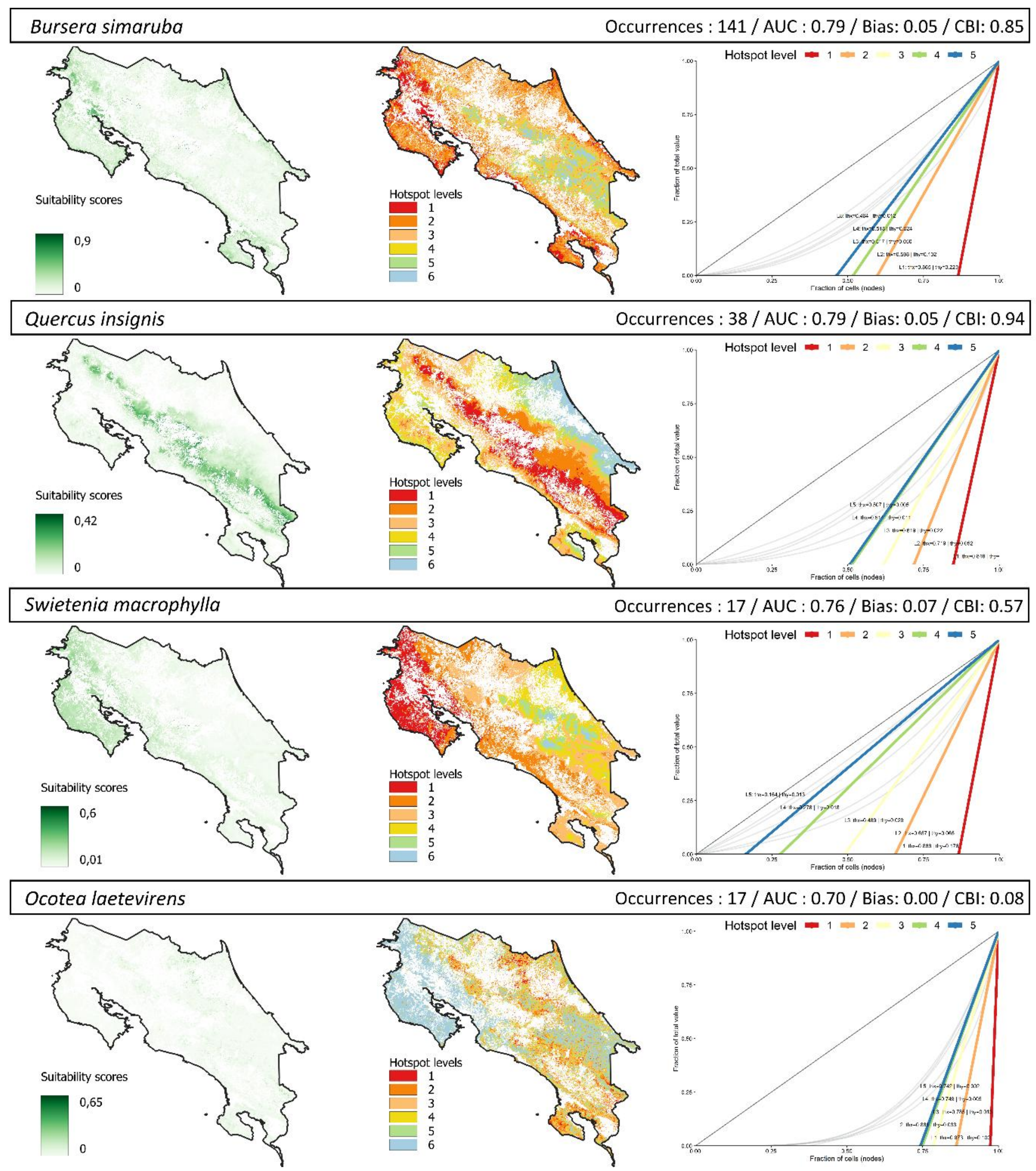

**Figure 3.** For four species (*Bursera simaruba, Quercus insignis, Swietenia macrophylla, Ocotea laetivirens*), this figure illustrates (i) the final habitat suitability maps produced by the ESDM (left panel), (ii) their discretisation into hotspot levels using the Loubar approach (central panel), and (iii) the corresponding Lorenz curves used to define spatial concentration thresholds (right panel). For each species, the number of occurrences used and model performance metrics (AUC, Bias, CBI) are reported in the header, spanning situations from high to very low sample sizes (down to 17 occurrences) and variable model performance.

The importance of variables varied considerably among species, with medians ranging from 7.87 % (NDWIw sd) to 14.5 % (altitude) (Figure 4). This variability was particularly pronounced for altitude, seasonal precipitation (PRSea), soil pH at 30 cm (pH30), and cation exchange capacity at 30 cm (CEC30). In contrast, some variables displayed more compact distributions (NDWIw mean, sd, anPR, and slope mean), indicating relatively consistent importance across species. Extreme values were observed for nearly all variables, except pH30, suggesting that some species assign much greater importance to certain variables than the majority. Altitude showed the highest extreme values, with some species attributing over 50–60 % importance, while others were around 5–10 %.

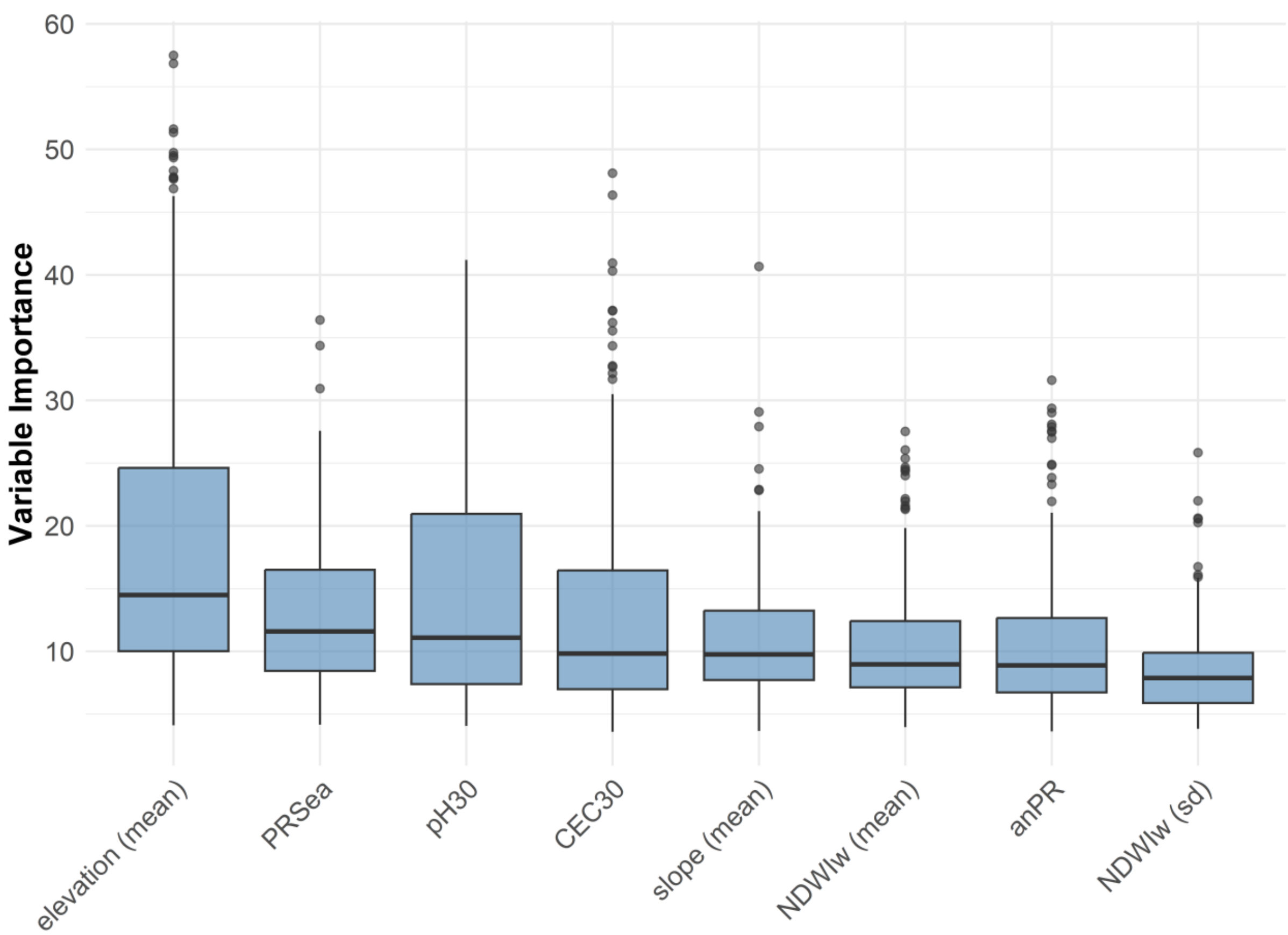

**Figure 4.** Distribution of the relative importance of the eight selected non-collinear environmental variables used as inputs in the SDMs.

### 3.2. Development of a local anthropogenic pressure indicator derived from LULC data

Cross-scale sensitivity analyses (pseudo-variogram, global variance, and inter-scale stability) converged toward a stabilisation of landscape-metric behaviour at 5 km (20 pixels), indicating limited gain in information beyond this scale and reduced noise relative to smaller windows. We therefore used a 5-km focal window for subsequent moving-window metric computations. Detailed results are provided in Figure S3. Landscape metrics at the class level were therefore computed using a 5 km focal window, to develop the anthropogenic pressure indices for the LULC classes Forest Plantation, Crop, Pasture, and Urban Area.

For these four anthropogenic classes, the metrics retained following the collinearity analysis ($|r| < 0.8$) were edge density (ED), number of patches (NP), the coefficient of variation of the fractal dimension index (FRAC_CV), and the minimum distance to pixels of the target class (MinDist) (Figure S4). All of them contributed significantly to the primary gradient (PC1) extracted by the class-specific PCA runs (Table 1). The first component (PC1) consistently explains over 70% of the variance, confirming the existence of a dominant gradient effectively summarising the structural variations specific to each class. The loadings associated with the metrics are all positive and of similar magnitude across classes, indicating that all descriptors change in the same direction along the pressure gradient. This convergence reflects a highly stable internal structure: the same landscape dimensions drive variability, regardless of the anthropogenic class considered. This coherence allows PC1 to be used directly as a synthetic pressure index for each class. The resulting anthropogenic pressure indices display marked contrasts across Costa Rica (Figure 5). Croplands (Figure 5b) and pastures (Figure 5c) emerge as the main drivers of pressure within forested landscapes nationwide. Forest plantations (Figure 5a) exhibit a patchier but locally intense distribution, particularly in the north-Caribbean regions. Urban areas (Figure 5d), though spatially limited, generate high-pressure levels around the Central Valley, where the capital is located, and around major secondary cities.

**Table 1.** Loadings of landscape metrics—edge density (ed), number of patches (np), coefficient of variation of the fractal dimension index (frac_cv), and minimum distance to the pixel of the target class (min dist)—on the first principal component (PC1) for different anthropogenic land-use classes (forest plantation, crop, pasture, urban area). The Direction indicates the expected relationship between each metric and the anthropization gradient (+: increase with anthropization; –: decrease).

| Class | ed | np | frac_cv | min dist | CP1 |
|---|---|---|---|---|---|
| Forest Plantation | 0.52 | 0.53 | 0.48 | 0.47 | 0.80 |
| Crop | 0.50 | 0.52 | 0.48 | 0.50 | 0.77 |
| Pasture | 0.49 | 0.49 | 0.49 | 0.52 | 0.73 |
| Urban area | 0.51 | 0.55 | 0.46 | 0.47 | 0.74 |
| Direction | + | + | + | – | |

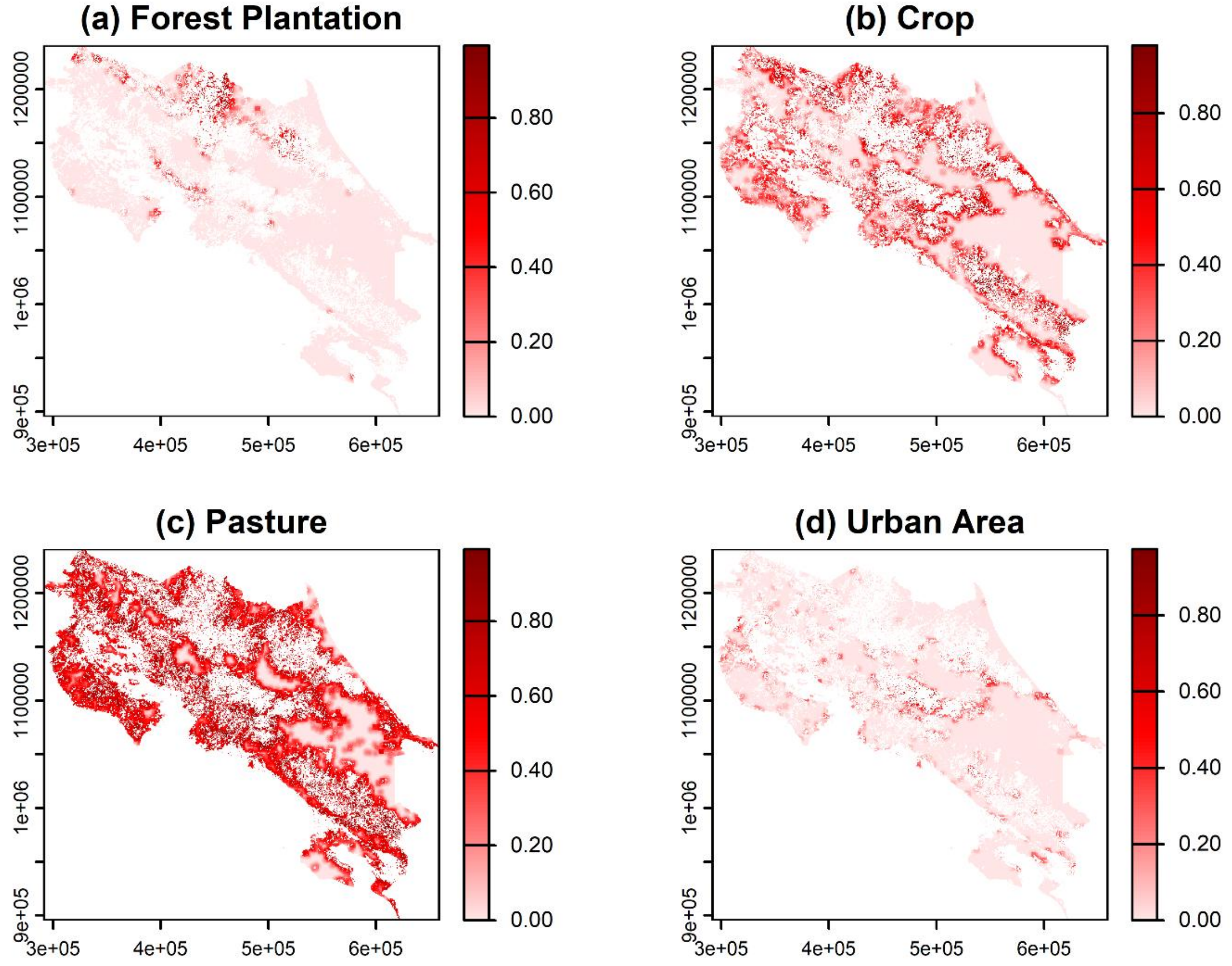

**Figure 5.** Anthropogenic pressure maps and distribution of forest species hotspots in Costa Rica. (a–d) Anthropogenic pressure indices derived from four LULC classes: forest plantations, croplands, pastures, and urban areas. Each index is normalized from 0 to 1 based on the same four landscape metrics: edge density (ed), number of patches (np), coefficient of variation of the fractal dimension index (frac_cv), and minimum distance to the pixel of the target class (min dist).

### 3.3. Mapping of conflict and refuge areas: Overlap between anthropogenic pressure and biodiversity hotspots

The overlap of hotspot levels as defined by the Loubar method for potential biodiversity $b_x$ and actual local anthropogenic pressure $p_x$ within Costa Rica's forest cover reveals significant spatial structuring (Figures 6 and 7).

High levels of potential biodiversity hotspots largely dominate the forest territory, with 47.6% in $b_1$ and 39.6% in $b_2$, mainly concentrated in mountainous forest complexes, the northern Caribbean coast, and the southern Pacific coast (Osa Peninsula), which may contain between 70% and 80% of species. In contrast, low levels remain marginal (< 0.5%), primarily in the Central and Talamanca mountain ranges, containing at least 8.66% of species (Figures 6a, 6c, S5). Anthropogenic pressure follows a comparable hierarchy but along a different geographical gradient. Its highest levels cover 37.7% ($p_1$) and 38% ($p_2$) of forests, especially in foothills, lowlands, and coastal areas across the country. Meanwhile, low levels ($p_4$ –$p_6$: 6.3%, 2.3%, 1.4%) are concentrated in large protected contiguous forest areas and more isolated areas (Figure 6b, c). Lorenz curves with the thresholds for each hotspot level of anthropogenic pressure and potential biodiversity are presented in Figure S6.

The associated continuous results $B(x)$ and $P(x)$, whose spatial representation is shown in Figure S5, confirm through a scatter plot a generally increasing but highly dispersed relationship between anthropogenic pressure and potential species richness. Projecting the discrete $(b, p)$ combinations into the continuous $(B, P)$ space highlights distinct clusters, which were used to define four operational categories a posteriori: high conflict, conflict, priority refuge, and refuge (Figure 6c; Table 2).

**Table 2.** Operational categories derived from combinations of biodiversity b and anthropogenic pressure p hotspot levels

| Priority area | Combinations of hotspot levels | Interpretation |
|---|---|---|
| High conflict | b1p1 | Maximum potential biodiversity and maximum anthropogenic pressure |
| Conflict | b1p2 , b2p1 | High-to-maximum potential biodiversity and very high anthropogenic pressure |
| Priority refuge | b1p4 , b1p5 , b1p6 | Maximum potential biodiversity and low anthropogenic pressure |
| Refuge | b2p4 , b2p5 , b2p6 | High potential biodiversity and low anthropogenic pressure |

Spatial integration of these combinations shows that high-conflict areas cover 23.42% of forest area and moderate-conflict areas 30.18%, mainly in lowlands. In contrast, priority refuges (1.76%) and refuges (4.96%) are restricted and concentrated in mountainous contiguous forest area, the Osa Peninsula, and certain northern Caribbean forests (Figure 7). Their extent departs markedly from a random distribution (p < 0.001; Z = 3–780), revealing a strong and statistically significant spatial association between potential biodiversity hotspots and anthropogenic pressure hotspots. Finally, refuges and priority refuges remain predominantly located within protected areas (only 7.26% and 8.38% outside, including 3.36% and 1.28% within biological corridors), whereas the majority of conflict and high-conflict zones lie outside these areas (89.83% and 69.47%), including 51% and 40.8% within corridors. This pattern highlights the vulnerability of forest sectors exposed to the highest levels of anthropogenic pressure. Figure 7 reveals a recurrent spatial organisation of Costa Rica's protected continuous forests, in which protected areas are frequently surrounded by a peripheral belt of conflict zones, followed by an intermediate transition zone, and finally a core dominated by refuge and priority refuge areas. This concentric configuration suggests that the ecological integrity of protected forest cores increasingly depends on the effective management and restoration of surrounding high-pressure landscapes.

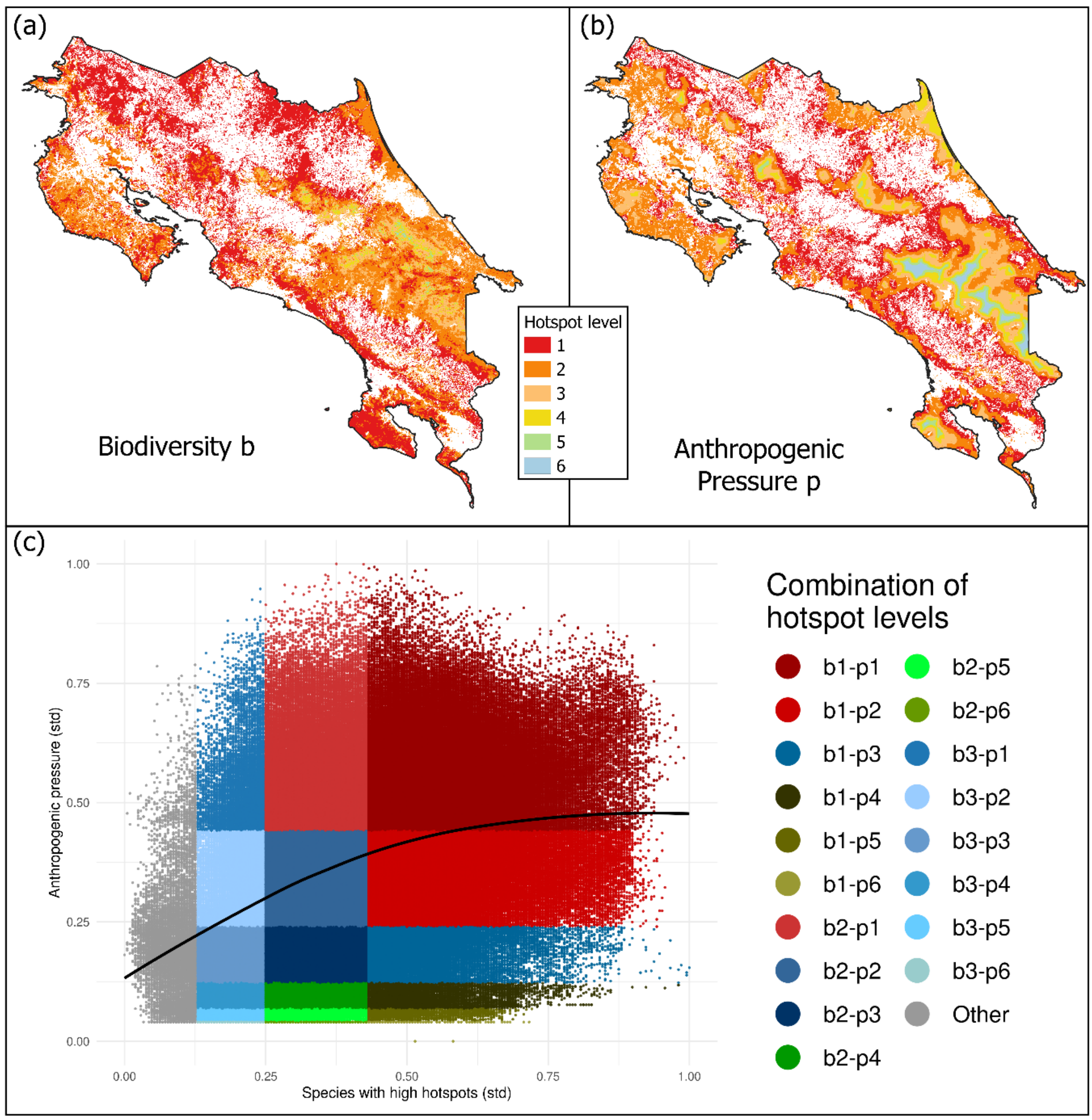

**Figure 6.** Overlap between potential biodiversity (b) and anthropogenic pressure (p) hotspots across the national forest cover. The hotspot maps (b) and (p), containing six loubar hotspot levels (1 = maximum intensity, 6 = minimum). The lower panel represents, for each forest pixel, the normalised intensity of anthropogenic pressure as a function of normalised species richness. Points are coloured according to their combination of hotspot levels (b–p), highlighting potential conflict

areas (red) and refuge areas (green). The loess curve illustrates the general trend between the two variables.

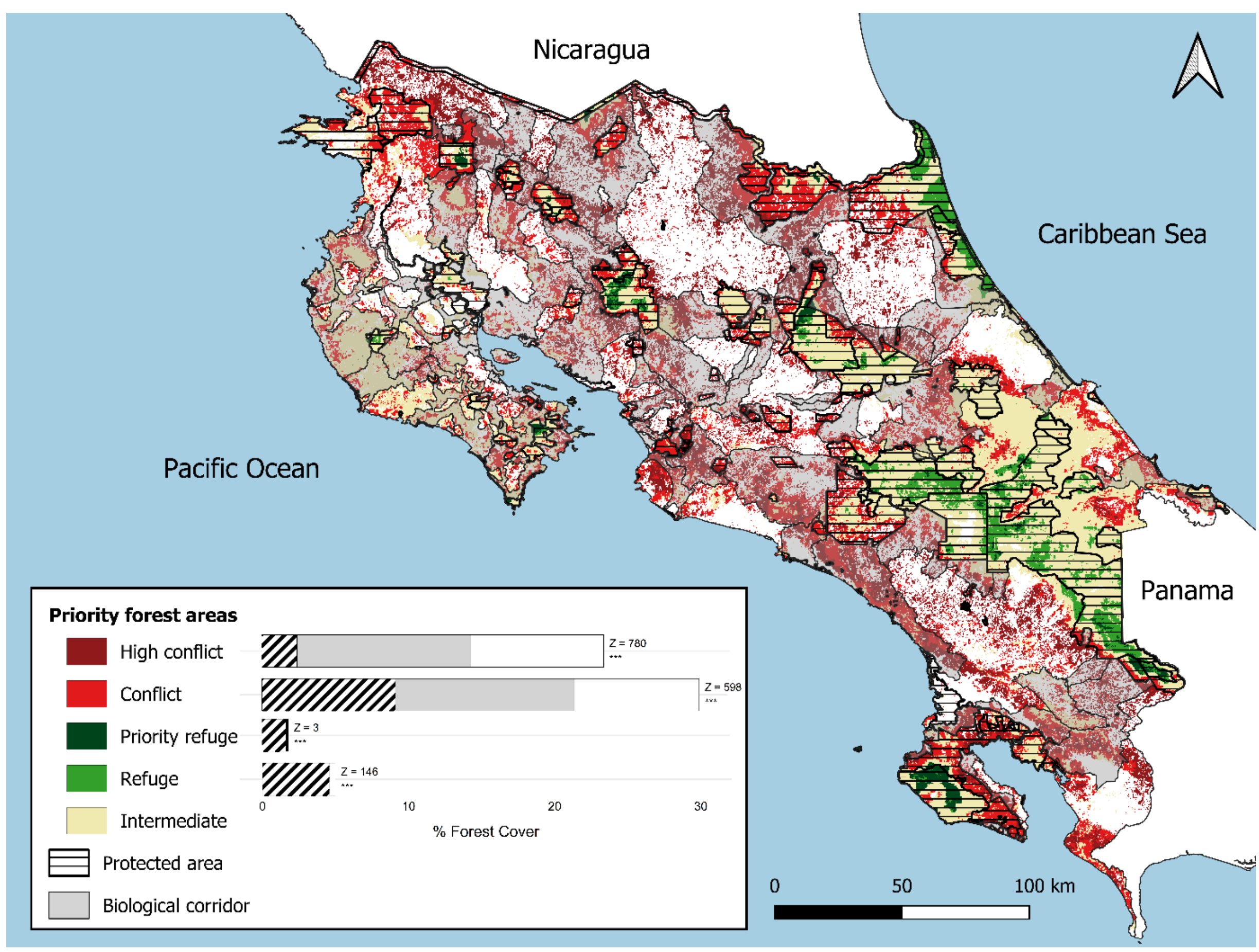

**Figure 7.** Map of priority forest areas in Costa Rica, identified by overlaying potential biodiversity hotspots of tree species that structure the main forest ecosystems with anthropogenic pressures at the landscape scale (5 km), with the addition of protected area boundaries. Five priority areas were distinguished: high-conflictconflict, priority biodiversity refuge, refuge, and intermediate. The adjacent bar plots show, for each category, the proportion of national forest cover, accompanied by its z-score and significance level (p < 0.001; ***), distinguishing surfaces within protected areas (hatched) from those outside (light grey).

## 4. Discussion

This study developed a reproducible, spatially explicit indicator for national biodiversity conservation planning. Designed from open-access biodiversity and land-use data, the indicator is intended to support implementation and reporting under the Kunming–Montreal Global Biodiversity Framework (GBF). It integrates potential ecological value, derived from multi-species distribution models of canopy-forming structural species used as an ecological proxy (Souza Oliveira et al., 2025a), with an index of exposure to anthropogenic pressures to identify priorities for protection, restoration, and sustainable management. Rather than aiming to provide an exhaustive representation of biodiversity, the indicator is conceived as a transparent and transferable decision-support tool that can inform spatial planning, conservation prioritisation, gap analyses of protected-area networks, and progress towards GBF Targets 1–3, which call for integrated spatial planning, ecosystem restoration, and representative, well-connected, and effectively managed conservation systems (Hughes and Grumbine, 2023; Ward et al., 2020).

### 4.1. Validation and interpretation of results in Costa Rica

At the scale of Costa Rica, potential refuges are concentrated within the major contiguous forest area(Figures 1, 7) already identified as priority forest cores, combining high species richness and high endemism, in national multi-taxon syntheses (Kohlmann et al., 2010) and in areas recognised as important for biodiversity (KBA) (CBD, 2021). These enclosed contiguous forest area are strongly protected and correspond to the Lowland wet evergreen forest of Caribbean slope in the Northeastern Caribbean Lowland (Barra del Colorado and Tortuguero), Wet seasonal evergreen forest on the Osa Peninsula and Premontane-to-mountain mixed-to-evergreen cloud forest in the Guanacaste, Tilarán and Central cordilleras, and Mountain oak rainforest of Talamanca (Souza Oliveira et al., 2025a). The only extensive priority refuge identified corresponds to the core of the Osa Peninsula (Corcovado National Park), confirming the importance of this region as a national hotspot of biodiversity and endemism, favoured by strong environmental and topographic variability (Gilbert and Kappelle, 2016; Hofhansl et al., 2019; Zamora et al., 2004). This strong convergence suggests that, at the national, i.e. macro-ecological, scale, despite the use of dominant canopy species as a proxy and despite biases and heterogeneity in occurrence data, the framework reproduces spatial gradients consistent with expected cores of high biodiversity value and ecological integrity (Cooper et al., 2024; Lima et al., 2026; Metcalfe et al., 2025; ter Steege et al., 2023).

Overlaying this indicator with Costa Rica's current conservation networks made it possible to identify recurrent spatial patterns within protected areas that display a high ecological value and low-pressure core (refuge), a highly exposed peripheral crown with strong ecological value (conflict), and an intermediate zone between these extremes (Figure 7). This pattern is consistent with global findings: a large proportion of protected forest areas are subject to increasing pressures and edge effects (microclimate, mortality, structural degradation), which strongly affect habitat quality, sometimes at considerable distances from the forest edge (Bourgoin et al., 2024; Grantham et al., 2020; Yang et al., 2025). In Costa Rica, a non-negligible share of conflict (10%) and high-conflict (20%) areas affects the margins of protected areas — and in some cases penetrates into them, particularly in the lowland contiguous forest area of dry deciduous forest in Guanacaste and wet evergreen forest of the Caribbean slope (San Juan–La Selva corridor), as well as part of the Osa Peninsula (Figure 1 – Souza Oliveira et al., 2025a). National assessments also acknowledge that degradation affects protected areas, due to persistent fragmentation linked in particular to the sale of usage rights ("venta de derechos") authorising agricultural, forestry or construction activities within Conservation Areas, thereby weakening the integrity of protected forests and potentially reducing their effectiveness, which may explain the observed patterns (MINAE, 2015).

The development of biological corridors is an essential strategy for maintaining connectivity among protected areas (MINAE, 2015; Morera Beita et al., 2021; Salom-Pérez et al., 2021). However, our indicator reveals that most conflict and high-conflict areas (90% and 70%, respectively) occur outside protected areas, with more than half overlapping with biological corridors. This finding highlights a potential vulnerability of the national connectivity network, as many corridor areas remain exposed to substantial anthropogenic pressures. These areas are frequently fragmented, predominantly located on private lands, and may not provide sufficiently suitable conditions to maintain the ecological functions expected from connectivity pathways for structuring species. Furthermore, existing corridors do not systematically incorporate altitudinal gradients, which are increasingly recognised as essential to facilitate species movements and climate-driven range shifts in tropical landscapes (Fung et al., 2017; Morera Beita et al., 2021). Moreover, these corridors do not always incorporate altitudinal gradients, which are essential for facilitating vertical species movements and climate-driven range shifts (Fung et al., 2017; Morera Beita et al., 2021). High-conflict areas are concentrated in fragmented forest patches embedded within agricultural lowland matrices and in some lowland–premontane forest complexes (e.g., Fila Costeña and foothill landscapes), representing multi-ecosystem areas where a high diversity of suitable niches coexist (Figures 1 and 7; Souza Oliveira et al., 2025a). Our results therefore reveal a spatial disconnect between highly anthropised lowland landscapes, where ecological connectivity is under pressure, and residual forest refuges that are increasingly concentrated in higher-elevation forest complexes. . Low-altitude protected areas therefore appear fragmented and isolated by deforestation and agricultural expansion (McCullough et al., 2024). This configuration reflects both the legacy of agricultural colonisation and the role of topographic constraints, which have limited land-use expansion in mountainous areas, highlighting the need to strengthen connectivity and protection in low-pressure zones (Redo et al., 2012; Shaver et al., 2015). Altitude emerges as a major determinant of the distribution of many species, increasing vulnerability to warming and shifting isotherms (Aguirre-Gutiérrez et al., 2025a). In this context, cordillera refuges constitute an asset for mountain ecosystems, which are highly endemic and spatially confined. Conversely, the dry lowland forests of the north Pacific (Nicoya–Guanacaste) combine environmental constraints (altitude, PRSea) and isolation linked to agricultural expansion, increasing their sensitivity to droughts, fires and anthropogenic disturbances (Figure 1 – Souza Oliveira et al., 2025a). This issue of isolation goes beyond the case of Costa Rica: global indicators show that protected area connectivity remains a frequent shortcoming with regard to the objective of well-connected networks (Saura et al., 2018).

This interpretation provides a transferable insight for other tropical countries: the effectiveness of protected-area networks depends not only on the ecological value conserved within their boundaries, but also on the intensity of surrounding pressure gradients and the maintenance of functional connectivity across the wider landscape matrix.

### 4.2. Translating results into protection and restoration strategies

Within the framework of the Kunming–Montreal Global Biodiversity, which calls for integrated approaches to biodiversity protection, ecosystem restoration, and effective management by 2030 (CBD, 2022), our refuge–conflict indicator offers a reproducible decision-support tool for identifying spatially explicit conservation priorities at the national scale in tropical forest ecosystems (Figure 7).

Areas categorised as “refuges” correspond to spaces combining high potential ecological value and low anthropogenic pressure. In these sectors, the priority is to protect the ecological integrity of these core areas, followed by their spatial expansion. The literature shows that protected status does not necessarily guarantee a high level of integrity (Grantham et al., 2020; Shrestha et al., 2021). In the Costa Rican context, this implies preventing internal fragmentation, limiting authorised land-use changes within Conservation Areas, strengthening monitoring capacities, and explicitly integrating indicators of ecological value and ecological integrity into management plans (Dias et al., 2023; MINAE, 2015). Maintaining connectivity between protected areas and along altitudinal gradients also appears crucial in order to preserve species’ movement options in response to shifting isotherms (Baumbach et al., 2021; McCullough et al., 2024).

Conversely, “conflict” areas, characterised by high potential ecological value coupled with strong anthropogenic pressures, designate zones where restoration is a priority in order to promote ecological integrity and regeneration, thereby making biodiversity potential effective. In Costa Rica, these areas are mainly concentrated at the margins of protected areas, within biological corridors on private land, and in fragmented agricultural mosaics of the lowlands or premontane zones (notably Guanacaste, the Caribbean foothills and the Fila Costeña). In these contexts, the priority lies in functional restoration aimed at improving matrix permeability and reducing edge effects (Bourgoin et al., 2024; Nunes et al., 2023; Yang et al., 2025). This implies a hierarchical restoration strategy: (i) promote natural regeneration where ecologically plausible (passive restoration); (ii) deploy active restoration only where the matrix is too degraded to allow the spontaneous return of biodiversity (planting/enrichment with native and resilient species); (iii) improve matrix permeability through agroforestry; and (iv) target incentives (PES) at corridor sections acting as bottlenecks (Chazdon et al., 2025; Cole et al., 2024; Holl et al., 2022). This interpretation is consistent with studies highlighting that biodiversity–carbon–cost restoration benefits vary greatly depending on location and cannot be optimised through simple quantitative targets (Strassburg et al., 2020).

Particular attention must be paid to interfaces between refuges and conflict areas. Where protected areas are bordered by high-pressure sectors, peripheral restoration can stabilise the intermediate zone identified in our analysis by reducing the intensity of edge effects and increasing core habitat area. This phenomenon is particularly relevant in the dry lowlands of the north Pacific, the San Juan–La Selva corridor on the northern Caribbean coast, and on the Osa Peninsula, where external fragmentation may compromise the internal ecological functionality of protected cores. The multi-species mapping produced at 250 m resolution thus makes it possible to guide, in a spatially explicit manner, action priorities across the 60% of forest cover identified as priority for the conservation of the species studied. Beyond the Costa Rican case, these results recall that the performance of a protected area network does not depend solely on the protected zone itself, but also on pressure gradients at its margins and the functional continuity of the surrounding matrix (Brennan et al., 2022; Jones et al., 2018). The refuge–conflict approach thus provides a spatially explicit workflow for moving from area-based targets to planning grounded in ecological potential, combined with integrity, connectivity and the strategic reduction of pressures.

**4.3. Limitations and perspectives**

Dominant canopy tree species structuring ecosystems provide a biodiversity proxy consistent with forest structure and allow identification of a significant share of potentially priority areas at national scale. However, this non-exhaustive approach may under-represent key components of biodiversity (rare, endemic, specialist, understorey species, fauna) and thus underestimate the representativeness of spatial priorities (Aguirre-Gutiérrez et al., 2025b; Barlow et al., 2007; Ritter et al., 2019; Vallé et al., 2025). Consequently, refuge–conflict maps should be viewed as a tool for identifying priority areas rather than as a complete and definitive mapping of reality. Subsequently, it will be important to test the robustness of these priority areas through an extension of the proxy to multi-taxon and/or endemic and threatened species.

This bias is amplified by the use of GBIF data, which largely rely on citizen science databases and are concentrated in accessible areas, leaving under-sampled hard-to-access regions such as the Talamanca cordillera or the northern Caribbean coastal forests, which nevertheless belong to the richest and most endemic regions. The integration of complementary data from in situ campaigns, systematic surveys or fine-scale remote sensing technologies could reduce this bias and improve the spatial accuracy of refuges (Fernández et al., 2020). However, data acquisition remains a major obstacle: high costs, lack of standardisation and coordination difficulties lead to strong disparities in in situ data, limiting countries' capacity to monitor and model biodiversity (Chapman et al., 2024).

An anthropogenic pressure index based on LULC classes and weightings effectively captures local gradients of artificialisation (deforestation, urbanisation) and may indirectly reflect certain pressures (pollution, human density) (Grantham et al., 2020). Nevertheless, it may under-represent major threats that are less visible in LULC (hunting, selective logging, fires, land-use conflicts, invasive species), and suffer from temporal mismatches between biological occurrences, infrastructure and LULC mapping. Furthermore, a purely "local" reading struggles to capture the spatio-temporal cumulative nature of direct and indirect pressures, which is nevertheless crucial for assessing vulnerability (Martini et al., 2024). It would be of interest to couple LULC-based pressure with integrity layers and more mechanistic pressure indicators; in this regard, the FLII provides a useful reference integrating observed/inferred pressures and large-scale connectivity loss (Grantham et al., 2020). This workflow also aims to incorporate climate-change scenarios, which are essential for adaptive and forward-looking planning (Keppel et al., 2015, 2012; Selwood and Zimmer, 2020).

An often implicit limitation is that ecological prioritisation does not guarantee feasibility (opportunity costs, land tenure, governance, acceptability). Thus, the refuge–conflict indicator should be considered as a synthetic module combining potential ecological value and relative exposure to anthropogenic pressures, which can be integrated into national planning processes. It does not capture all the determining dimensions for tropical forest conservation, such as opportunity costs, land tenure and social constraints, governance capacities, or specific threats (e.g. fires, hunting and invasive species). Nevertheless, it makes explicit a central trade-off in conservation, by identifying areas where ecological potential is high but likely to be rapidly compromised by local anthropogenic pressures. Hence the relevance of this indicator in a context where expected benefits (biodiversity, carbon, costs) vary greatly depending on location and the weighting of multiple objectives (Strassburg et al., 2020). This indicator could therefore be integrated into a Systematic Conservation Planning (SCP – Plumptre et al., 2025) framework to inform multi-action zoning (protection, restoration, management, connectivity) (Baker et al., 2025; Hanson et al., 2025) with explicit connectivity indicators (Saura et al., 2018) and, where objectives are heterogeneous, support multi-criteria decision analysis approaches that make assumptions and trade-offs explicit (Cavalcante et al., 2022). In the Costa Rican case, operationalisation could be facilitated by the existence of already structured national frameworks, GRUAS and the National Biological Corridors Programme (PNCB–SINAC), which provide an institutional foundation for translating a synthetic indicator into targeting rules, scenarios and action priorities (SINAC, n.d.).

## Conclusion

Ultimately, the refuge–conflict indicator should not be interpreted as a prescriptive map defining where conservation actions must occur, but rather as a transparent decision-support tool that helps reveal spatial trade-offs in tropical forest planning. By integrating ecological potential with exposure to anthropogenic pressures, it moves beyond static biodiversity mapping towards a more dynamic, risk-informed approach for prioritising conservation investments across protection, restoration, and management actions. This framework with the proposed workflow provides a reproducible and adaptable basis for supporting national biodiversity strategies, including the identification of priorities relevant to the implementation and reporting of global biodiversity targets under increasing fiscal, social, and climatic constraints.

Although its application depends on governance capacity, stakeholder engagement, and complementary socio-economic assessments, the integration of such synthetic spatial indicators into national planning processes can improve the alignment between conservation objectives and on-the-ground implementation. As tropical forests continue to face accelerating pressures and countries seek to meet ambitious biodiversity and climate commitments, approaches that make ecological priorities, underlying assumptions, and spatial trade-offs explicit will be essential for designing effective and realistic conservation strategies.

## CRediT authorship contribution statement

Maïri Souza Oliveira: Conceptualization, , Data curation, Methodology, Formal analysis, Investigation, Validation, Writing – original draft, Writing – review and editing.
Maxime Lenormand: Conceptualization, Methodology, Investigation, Writing – review, Supervision.
Sandra Luque: Conceptualization, Methodology, Investigation, Writing – review and editing, Funding acquisition, Project administration, Supervision.

## Acknowledgements

This research was partially funded by a PhD fellowship from INRAE (National Research Institute on Agriculture, Food & the Environment) and CIRAD (Centre for International Cooperation in Agricultural Research for Development), with additional support from CNES (National Space Agency, France) TOSCA Project CEOS-CODEX for fieldwork and data validation. We are grateful to Mona Bonier, Florian de Boissieu, Rémy Découpe, Cassio Fraga Dantas, Jean-Baptiste Féret, Renaud Marti and Simon Madec (TETIS research unit) for their valuable support in various aspects of data analysis and expertise. Thanks also to Adriana Aguilar Porras (SINAC - Sistema Nacional de Áreas de Conservación), Nelson Zamora (ITCR - Instituto Tecnológico de Costa Rica) and Diego Delgado (CATIE - Centro Agronómico Tropical de Investigación y Enseñanza) for their field expertise.

## Declaration of competing interest

The authors declare that they have no known competing financial interests or personal relationships that could have appeared to influence the work reported in this paper.

## Data Availability Statement

The original occurrence data were obtained from the Global Biodiversity Information Facility (GBIF) and are cited in the main text and reference list. The final dataset used in the analyses, the filtering decisions, and the processing and analysis scripts are available in the Git repository at https://github.com/ mairi-souzaoliveira /CR_refuge-conflict_areas.

## Supplementary material

**Section S1.** Cross-scale sensitivity analysis to select the moving-window size

Landscape metrics derived from categorical LULC maps are scale dependent, and moving-window ("focal") calculations are commonly used to represent spatial variation in landscape composition/configuration as continuous surfaces. Because the window size controls the observation scale, we performed a cross-scale sensitivity analysis across a set of candidate window sizes and retained the window that provided a good compromise between (i) preservation of local spatial variability, (ii) avoidance of excessive smoothing, and (iii) consistency across scales (Hagen-Zanker, 2016).

Let $m_w(x)$ denote the raster value of a given class-level landscape metric at pixel location $x$, computed using a square moving window of side length $w$ pixels (odd integers). Let $s$=250m, the pixel size. Candidate windows were $w \in \{5,11,21,31,41\}$ pixels. For each metric $m$ and each window size $w$, we computed three complementary indices: (i) spatial roughness, (ii) global variance, and (iii) inter-scale stability.

(i) Spatial roughness index (mean absolute central difference; "MACD roughness") – To quantify how rapidly the metric surface varies locally (and how this variation decreases as the window size increases), we computed a roughness index based on absolute central differences in the horizontal and vertical directions. For raster cell$(i, j)$, define:

$\Delta_x(i,j) = |m_w(i,j+1) - m_w(i,j-1)|$,

$\Delta_y(i,j) = |m_w(i+1,j) - m_w(i-1,j)|$.

The spatial roughness index is then:

$$R_w = \frac{1}{2}(\Delta_x + \Delta_y)$$

where the overbar denotes the mean over all cells for which the required neighbours exist and are non-missing.

(ii) Global variance – Global variance was computed as the (non-missing) variance of raster cell values of $m_w$, to characterise how overall map variability changes with window size (Šímová and Gdulová, 2012).

(iii) Inter-scale stability – Inter-scale stability was measured as the Pearson correlation between metric surfaces obtained at two window sizes $w$, evaluated over the intersection of pixels with valid values in both rasters.

**Table S1.** Exploration of the occurrence data of the studied species: "GBIF abundance" representing the number of occurrences of the species in the study area included in the GBIF data and, "Abundance (res 250m)" indicating species abundance at a spatial resolution of 250 m with at least 10 occurrences.

| Species | GBIF abundance | Abundance (res 250m) |
|---|---|---|
| Adelia triloba | 42 | 13 |
| Aegiphila anomala | 10 | 0 |
| Aiouea montana | 12 | 0 |
| Albizia adinocephala | 51 | 23 |
| Albizia niopoides | 24 | 0 |
| Alchornea costaricensis | 29 | 11 |
| Alchornea glandulosa | 34 | 0 |
| Alchornea latifolia | 66 | 19 |
| Alchorneopsis floribunda | 20 | 0 |
| Alfaroa costaricensis | 57 | 19 |
| Alnus acuminata | 43 | 23 |
| Amyris pinnata | 40 | 10 |
| Anacardium excelsum | 77 | 46 |
| Anaxagorea crassipetala | 46 | 15 |
| Andira inermis | 61 | 31 |
| Annona montana | 11 | 0 |
| Annona papilionella | 14 | 0 |
| Annona purpurea | 26 | 17 |
| Apeiba membranacea | 86 | 58 |
| Apeiba tibourbou | 138 | 98 |
| Aralia excelsa | 27 | 0 |
| Ardisia compressa | 115 | 48 |
| Ardisia palmana | 38 | 15 |
| Ardisia pleurobotrya | 41 | 0 |
| Ardisia revoluta | 53 | 26 |
| Astronium graveolens | 34 | 17 |
| Axinaea costaricensis | 11 | 0 |
| Batocarpus costaricensis | 17 | 0 |
| Beilschmiedia costaricensis | 14 | 0 |
| Billia rosea | 145 | 45 |
| Bravaisia integerrima | 22 | 12 |

| Species | GBIF abundance | Abundance (res 250m) |
|---|---|---|
| Brosimum alicastrum | 69 | 34 |
| Brosimum guianense | 27 | 13 |
| Brosimum lactescens | 22 | 10 |
| Brosimum utile | 32 | 20 |
| Brunellia costaricensis | 46 | 15 |
| Bursera simaruba | 193 | 141 |
| Bursera tomentosa | 27 | 15 |
| Byrsonima arthropoda | 12 | 0 |
| Byrsonima crassifolia | 136 | 82 |
| Calatola costaricensis | 93 | 31 |
| Calliandra trinervia | 33 | 0 |
| Calophyllum brasiliense | 62 | 27 |
| Calycophyllum candidissimum | 64 | 38 |
| Carapa guianensis | 29 | 18 |
| Caryocar costaricense | 26 | 19 |
| Casearia arborea | 52 | 17 |
| Casearia corymbosa | 107 | 36 |
| Casearia sylvestris | 92 | 34 |
| Casearia tacanensis | 61 | 18 |
| Cassia grandis | 67 | 39 |
| Cassipourea elliptica | 104 | 28 |
| Castilla elastica | 55 | 26 |
| Castilla tunu | 30 | 15 |
| Cecropia angustifolia | 68 | 36 |
| Cecropia insignis | 62 | 44 |
| Cecropia obtusifolia | 189 | 151 |
| Cecropia peltata | 130 | 107 |
| Cedrela tonduzii | 22 | 14 |
| Celtis schippii | 12 | 0 |
| Cenostigma eriostachys | 20 | 12 |
| Chimarrhis latifolia | 32 | 14 |
| Chimarrhis parviflora | 35 | 13 |
| Chionanthus panamensis | 10 | 0 |
| Chomelia microloba | 20 | 0 |
| Chomelia spinosa | 64 | 21 |
| Chromolucuma congestifolia | 12 | 0 |
| Chrysochlamys allenii | 64 | 20 |
| Chrysochlamys glauca | 142 | 39 |
| Chrysophyllum brenesii | 69 | 19 |
| Citharexylum costaricense | 10 | 0 |
| Clarisia biflora | 57 | 21 |
| Clarisia racemosa | 10 | 0 |
| Cleidion castaneifolium | 39 | 0 |
| Clethra consimilis | 24 | 0 |
| Clethra costaricensis | 74 | 26 |

| Species | GBIF abundance | Abundance (res 250m) |
|---|---|---|
| Clethra lanata | 55 | 20 |
| Cleyera theaeoides | 66 | 18 |
| Coccoloba tuerckheimii | 51 | 16 |
| Cochlospermum vitifolium | 144 | 92 |
| Colubrina spinosa | 62 | 22 |
| Compsoneura excelsa | 54 | 18 |
| Conceveiba pleiostemona | 31 | 10 |
| Cordia alliodora | 54 | 27 |
| Cordia bicolor | 30 | 0 |
| Cordia cymosa | 59 | 25 |
| Cordia megalantha | 15 | 0 |
| Cordia panamensis | 39 | 14 |
| Cornutia pyramidata | 99 | 43 |
| Couma macrocarpa | 10 | 0 |
| Couratari guianensis | 21 | 0 |
| Croton draco | 63 | 36 |
| Croton schiedeanus | 100 | 41 |
| Croton smithianus | 14 | 0 |
| Cupania glabra | 38 | 15 |
| Cupania guatemalensis | 45 | 15 |
| Cupania juglandifolia | 15 | 0 |
| Cynometra retusa | 32 | 11 |
| Damburneya smithii | 19 | 0 |
| Dendropanax arboreus | 84 | 26 |
| Dendropanax caucanus | 11 | 0 |
| Dendropanax globosus | 36 | 11 |
| Dialium guianense | 33 | 14 |
| Diospyros juruensis | 16 | 0 |
| Diphysa americana | 27 | 18 |
| Dipteryx panamensis | 29 | 20 |
| Drimys granadensis | 134 | 62 |
| Drypetes brownii | 21 | 0 |
| Dussia macroprophyllata | 29 | 14 |
| Dystovomita paniculata | 25 | 0 |
| Elaeagia auriculata | 39 | 20 |
| Elaeagia glossostipula | 13 | 0 |
| Elaeoluma glabrescens | 13 | 0 |
| Enterolobium cyclocarpum | 108 | 86 |
| Erythrina poeppigiana | 35 | 29 |
| Escallonia myrtilloides | 63 | 30 |
| Eugenia hiraeifolia | 47 | 14 |
| Ferdinandusa panamensis | 40 | 13 |
| Ficus costaricana | 60 | 25 |
| Ficus crassiuscula | 18 | 12 |
| Ficus crocata | 19 | 0 |

| Species | GBIF abundance | Abundance (res 250m) |
|---|---|---|
| Ficus insipida | 42 | 31 |
| Ficus maxima | 31 | 15 |
| Ficus obtusifolia | 50 | 22 |
| Ficus tonduzii | 38 | 19 |
| Garcinia madruno | 19 | 12 |
| Gliricidia sepium | 69 | 49 |
| Goethalsia meiantha | 23 | 17 |
| Graffenrieda galeottii | 31 | 15 |
| Grias cauliflora | 27 | 16 |
| Guarea glabra | 192 | 50 |
| Guarea guidonia | 72 | 11 |
| Guarea kunthiana | 48 | 14 |
| Guarea rhopalocarpa | 97 | 26 |
| Guarea tonduzii | 36 | 0 |
| Guatteria aeruginosa | 88 | 24 |
| Guatteria oliviformis | 62 | 20 |
| Guatteria ucayalina | 13 | 0 |
| Guazuma ulmifolia | 109 | 63 |
| Guettarda macrosperma | 37 | 12 |
| Gymnanthes riparia | 41 | 10 |
| Gyrocarpus jatrophifolius | 28 | 0 |
| Hampea appendiculata | 92 | 34 |
| Handroanthus ochraceus | 48 | 33 |
| Hasseltia floribunda | 19 | 10 |
| Hasseltia guatemalensis | 45 | 0 |
| Hauya elegans | 25 | 0 |
| Hedyosmum bonplandianum | 129 | 40 |
| Hedyosmum goudotianum | 47 | 15 |
| Heisteria concinna | 33 | 17 |
| Heliocarpus americanus | 24 | 0 |
| Heliocarpus appendiculatus | 37 | 12 |
| Hernandia didymantha | 12 | 0 |
| Hieronyma alchorneoides | 28 | 14 |
| Hieronyma oblonga | 119 | 28 |
| Hirtella triandra | 83 | 24 |
| Hura crepitans | 86 | 61 |
| Hymenaea courbaril | 55 | 31 |
| Ilex fortunensis | 27 | 0 |
| Ilex lamprophylla | 31 | 14 |
| Ilex pallida | 39 | 0 |
| Inga barbourii | 15 | 0 |
| Inga laevigata | 35 | 0 |
| Inga leonis | 19 | 0 |
| Inga marginata | 74 | 24 |
| Inga mortoniana | 16 | 0 |

| Species | GBIF abundance | Abundance (res 250m) |
|---|---|---|
| Inga oerstediana | 49 | 25 |
| Inga punctata | 109 | 31 |
| Inga sapindoides | 65 | 24 |
| Inga thibaudiana | 39 | 12 |
| Inga venusta | 26 | 10 |
| Inga vera | 88 | 44 |
| Jacaranda copaia | 15 | 12 |
| Jacaratia dolichaula | 40 | 20 |
| Karwinskia calderonii | 18 | 0 |
| Lacistema aggregatum | 94 | 34 |
| Ladenbergia brenesii | 36 | 10 |
| Laplacea fructicosa | 14 | 0 |
| Leptolobium panamense | 26 | 12 |
| Licaria misantlae | 23 | 0 |
| Lippia myriocephala | 10 | 0 |
| Lonchocarpus acuminatus | 25 | 0 |
| Lonchocarpus costaricensis | 14 | 0 |
| Lonchocarpus felipei | 32 | 14 |
| Lonchocarpus ferrugineus | 18 | 0 |
| Lonchocarpus guatemalensis | 26 | 12 |
| Lonchocarpus macrophyllus | 23 | 0 |
| Lonchocarpus minimiflorus | 36 | 14 |
| Lonchocarpus monteviridis | 12 | 0 |
| Lonchocarpus parviflorus | 13 | 0 |
| Lonchocarpus rugosus | 45 | 14 |
| Luehea candida | 43 | 21 |
| Luehea seemannii | 47 | 21 |
| Luehea speciosa | 40 | 13 |
| Lunania mexicana | 38 | 11 |
| Lysiloma auritum | 15 | 0 |
| Lysiloma divaricatum | 33 | 14 |
| Mabea occidentalis | 43 | 18 |
| Maclura tinctoria | 55 | 27 |
| Macrohasseltia macroterantha | 31 | 12 |
| Magnolia poasana | 28 | 10 |
| Manilkara chicle | 30 | 12 |
| Manilkara staminodella | 14 | 0 |
| Maranthes panamensis | 11 | 0 |
| Marila laxiflora | 17 | 0 |
| Matisia pacifica | 19 | 0 |
| Matudaea trinervia | 26 | 0 |
| Meliosma allenii | 22 | 0 |
| Meliosma glabrata | 91 | 21 |
| Meliosma idiopoda | 92 | 21 |
| Meliosma vernicosa | 16 | 0 |

| Species | GBIF abundance | Abundance (res 250m) |
|---|---|---|
| Mespilodaphne veraguensis | 79 | 29 |
| Metteniusa tessmanniana | 31 | 0 |
| Miconia affinis | 146 | 46 |
| Miconia bigibbosa | 14 | 0 |
| Miconia biperulifera | 33 | 12 |
| Miconia brenesii | 74 | 15 |
| Miconia brevitheca | 17 | 0 |
| Miconia conomicrantha | 102 | 33 |
| Miconia conorufescens | 70 | 26 |
| Miconia donaeana | 60 | 17 |
| Miconia durandii | 76 | 17 |
| Miconia elata | 45 | 13 |
| Miconia lasiopoda | 52 | 20 |
| Miconia multispicata | 16 | 0 |
| Miconia prasina | 95 | 22 |
| Miconia schnellii | 37 | 10 |
| Miconia tonduzii | 165 | 52 |
| Minquartia guianensis | 29 | 27 |
| Morisonia pittieri | 18 | 0 |
| Mortoniodendron anisophyllum | 92 | 18 |
| Myrcia chytraculia | 13 | 0 |
| Myrcia splendens | 174 | 40 |
| Myrcianthes fragrans | 48 | 13 |
| Myriocarpa longipes | 106 | 50 |
| Myrospermum frutescens | 18 | 0 |
| Myrsine coriacea | 139 | 45 |
| Nectandra membranacea | 44 | 15 |
| Nectandra reticulata | 19 | 0 |
| Ocotea atirrensis | 109 | 34 |
| Ocotea austinii | 18 | 0 |
| Ocotea insularis | 82 | 31 |
| Ocotea laetevirens | 42 | 17 |
| Ocotea macrophylla | 39 | 10 |
| Ocotea stenoneura | 17 | 0 |
| Oreomunnea mexicana | 19 | 0 |
| Oreomunnea pterocarpa | 26 | 14 |
| Oreopanax xalapensis | 60 | 23 |
| Otoba novogranatensis | 43 | 21 |
| Pachira aquatica | 114 | 63 |
| Panopsis costaricensis | 45 | 16 |
| Peltogyne purpurea | 15 | 0 |
| Pentaclethra macroloba | 110 | 68 |
| Pentagonia costaricensis | 27 | 0 |
| Pera arborea | 20 | 0 |
| Perebea hispidula | 47 | 13 |

| Species | GBIF abundance | Abundance (res 250m) |
|---|---|---|
| Perrottetia longistylis | 38 | 11 |
| Persea americana | 102 | 82 |
| Persea caerulea | 11 | 0 |
| Persea schiedeana | 24 | 16 |
| Phyllanthus skutchii | 18 | 0 |
| Piptocoma discolor | 18 | 0 |
| Piscidia carthagenensis | 14 | 0 |
| Platymiscium curuense | 20 | 0 |
| Pleuranthodendron lindenii | 24 | 11 |
| Pleurothyrium palmanum | 11 | 0 |
| Pochota fendleri | 125 | 87 |
| Podocarpus oleifolius | 49 | 15 |
| Posoqueria latifolia | 138 | 67 |
| Poulsenia armata | 19 | 10 |
| Pourouma bicolor | 45 | 26 |
| Pourouma minor | 31 | 15 |
| Pouteria laevigata | 12 | 0 |
| Pouteria torta | 32 | 15 |
| Protium confusum | 42 | 17 |
| Protium panamense | 12 | 0 |
| Protium pittieri | 27 | 15 |
| Protium ravenii | 42 | 19 |
| Prumnopitys standleyi | 40 | 22 |
| Pseudolmedia glabrata | 29 | 14 |
| Pseudosamanea guachapele | 24 | 10 |
| Psychotria berteroana | 111 | 41 |
| Pterocarpus michelianus | 30 | 0 |
| Pterocarpus officinalis | 22 | 14 |
| Pterocarpus rohrii | 22 | 12 |
| Quercus corrugata | 21 | 0 |
| Quercus costaricensis | 65 | 34 |
| Quercus insignis | 63 | 38 |
| Quercus oleoides | 50 | 27 |
| Quercus sapotifolia | 89 | 27 |
| Quetzalia occidentalis | 35 | 0 |
| Rehdera trinervis | 78 | 23 |
| Rinorea dasyadena | 35 | 12 |
| Ruagea glabra | 52 | 13 |
| Ruptiliocarpon caracolito | 41 | 16 |
| Sacoglottis trichogyna | 16 | 0 |
| Samanea saman | 70 | 45 |
| Sapium glandulosum | 53 | 20 |
| Saurauia montana | 184 | 59 |
| Schizolobium parahyba | 23 | 20 |
| Sciodaphyllum pittieri | 44 | 18 |

| Species | GBIF abundance | Abundance (res 250m) |
|---|---|---|
| Sebastiania pavoniana | 27 | 0 |
| Senna papillosa | 87 | 29 |
| Sideroxylon capiri | 59 | 26 |
| Sideroxylon portoricense | 10 | 0 |
| Simarouba amara | 31 | 22 |
| Simarouba glauca | 29 | 0 |
| Simira maxonii | 24 | 11 |
| Sloanea medusula | 20 | 0 |
| Sloanea terniflora | 72 | 23 |
| Sloanea zuliaensis | 27 | 0 |
| Solenandra mexicana | 18 | 0 |
| Sorocea affinis | 33 | 13 |
| Sorocea pubivena | 64 | 21 |
| Sorocea trophoides | 68 | 23 |
| Spondias mombin | 60 | 31 |
| Spondias purpurea | 52 | 35 |
| Staphylea occidentalis | 164 | 47 |
| Stephanopodium costaricense | 15 | 0 |
| Sterculia apetala | 46 | 26 |
| Stryphnodendron microstachyum | 31 | 14 |
| Styrax argenteus | 43 | 16 |
| Swartzia simplex | 46 | 25 |
| Swietenia macrophylla | 31 | 17 |
| Symphonia globulifera | 140 | 68 |
| Symplocos serrulata | 26 | 0 |
| Tabebuia rosea | 84 | 60 |
| Tabernaemontana donnell-smithii | 124 | 57 |
| Tabernaemontana glabra | 61 | 28 |
| Tabernaemontana longipes | 143 | 38 |
| Tachigali costaricensis | 28 | 0 |
| Talisia nervosa | 14 | 0 |
| Tapirira guianensis | 17 | 0 |
| Tapirira mexicana | 28 | 13 |
| Terminalia oblonga | 48 | 28 |
| Tetrathylacium johansenii | 13 | 0 |
| Tetrathylacium macrophyllum | 29 | 15 |
| Tetrorchidium euryphyllum | 71 | 17 |
| Thouinidium decandrum | 32 | 13 |
| Ticodendron incognitum | 74 | 37 |
| Trichilia americana | 48 | 18 |
| Trichilia martiana | 82 | 29 |
| Trichilia pleeana | 13 | 0 |
| Trichilia septentrionalis | 13 | 0 |
| Trichospermum galeottii | 60 | 16 |

| Species | GBIF abundance | Abundance (res 250m) |
|---|---|---|
| Triplaris melaenodendron | 57 | 22 |
| Trophis racemosa | 95 | 30 |
| Unonopsis pittieri | 17 | 0 |
| Verbesina oerstediana | 36 | 0 |
| Viburnum costaricanum | 126 | 47 |
| Virola koschnyi | 39 | 19 |
| Virola montana | 11 | 0 |
| Virola nobilis | 33 | 0 |
| Virola sebifera | 57 | 23 |
| Vismia baccifera | 131 | 57 |
| Vismia macrophylla | 49 | 24 |
| Vitex cooperi | 48 | 26 |
| Vochysia allenii | 11 | 0 |
| Vochysia ferruginea | 62 | 30 |
| Vochysia guatemalensis | 42 | 24 |
| Warszewiczia coccinea | 123 | 71 |
| Weinmannia pinnata | 77 | 18 |
| Xylopia sericea | 44 | 13 |
| Xylopia sericophylla | 14 | 0 |
| Zanthoxylum acuminatum | 56 | 16 |
| Zanthoxylum setulosum | 47 | 17 |
| Zinowiewia integerrima | 77 | 21 |

**Table S2.** Presentation of the fifteen class-level landscape metrics calculated for developing an anthropogenic pressure index for each class of interest.

| Metric | Name | Unit | Definition |
|---|---|---|---|
| minDist | Minimal distance | m | Measures the distance to the nearest patch of the class. |
| area_CV | Patch area (coefficient of variation) | % | Measures the area of the class's patches. |
| area_mean | Patch area (mean) | $m^2$ | |
| ca | Total class area | $m^2$ | Corresponds to the total area occupied by all class's patches in the landscape. |
| clumpy | Clumpiness index | _ | Measures the degree of spatial aggregation of a patch class relative to a random distribution. |
| cohesion | Patch cohesion index | % | Measures the degree of physical connectivity of a patch class within the landscape. |
| ed | Edge density | m/ha | Corresponds to the sum of the lengths of all edge segments involving patches of the class, divided by the total landscape area. |
| frac_cv | Fractal dimension index (cv) | % | Corresponds to the shape-complexity index of the class's patches. |
| frac_mean | Fractal dimension index (mean) | _ | |
| lsi | Landscape shape index | _ | Standardised measure of the total edge length of patch class in the landscape, adjusted by the total landscape area. |
| mesh | Effective mesh size | $m^2$ | Measures the effective mean patch size of a class, weighted by patch area and relative to the total landscape area. |
| np | Number of patches | _ | Corresponds to the number of patches of the given class in the landscape, reflecting the degree of subdivision or fragmentation of that class. |
| pland | % of Landscape | % | Percentage of the total landscape area occupied by class's patches. |
| shape_cv | Shape index (cv) | % | Measures the shape complexity of the patches within the class. |
| shape_mean | Shape index (mean) | _ | |

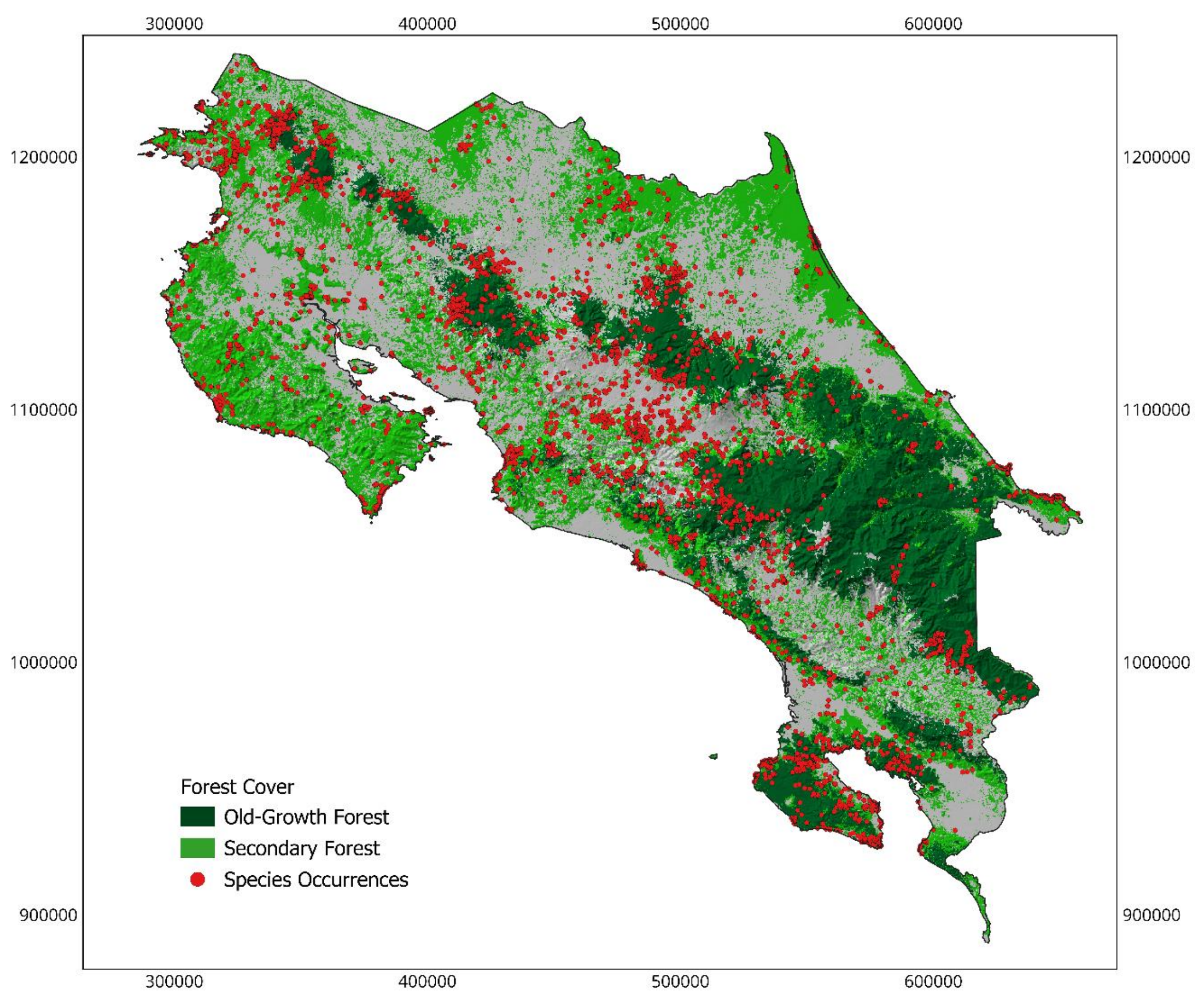

**Figure S1.** Distribution of occurrences of the 367 species studied from GBIF across Costa Rica's forest cover.

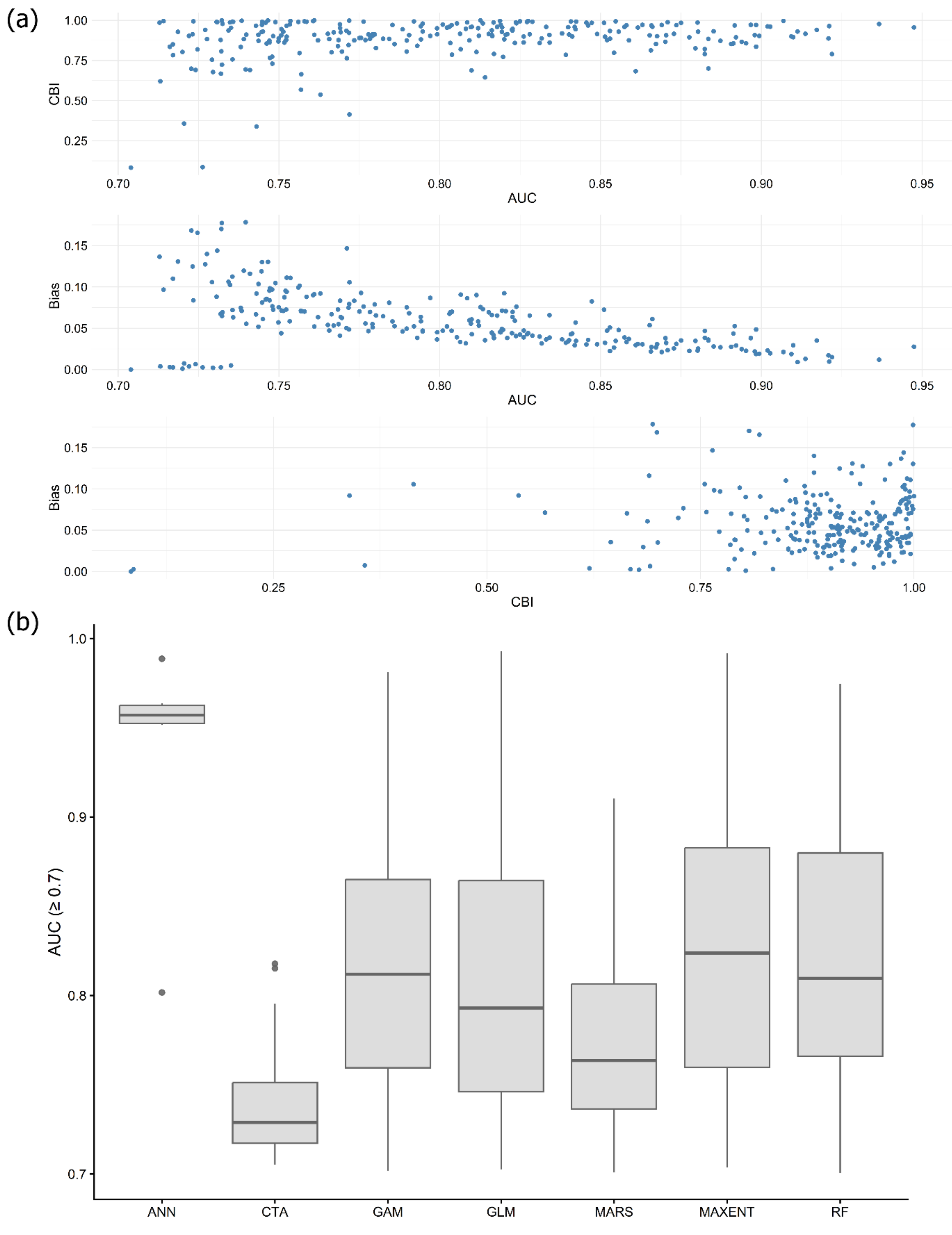

**Figure S2.** Performance evaluation of ESDMs: (a) Relationships between the performance evaluation metrics of the SDMs for the 254 modelled species: AUC, CBI, and bias. (b) Histogram of the seven modelling algorithms across all species based on AUC values for their inclusion in ensemble models (AUC ≥ 0.7): Generalised Additive Model (GAM), Generalised Linear Model (GLM), Multivariate Adaptive Regression Splines (MARS), Artificial Neural Networks (ANN), Classification Tree Analysis (CTA), Random Forest (RF), and Maximum Entropy (MAXENT). The ANN and CTA algorithms are the least frequently retained in the final ensemble: ANN is retained for only one species and CTA for 32 species, whereas the other algorithms are retained for between 213 and 247 species.

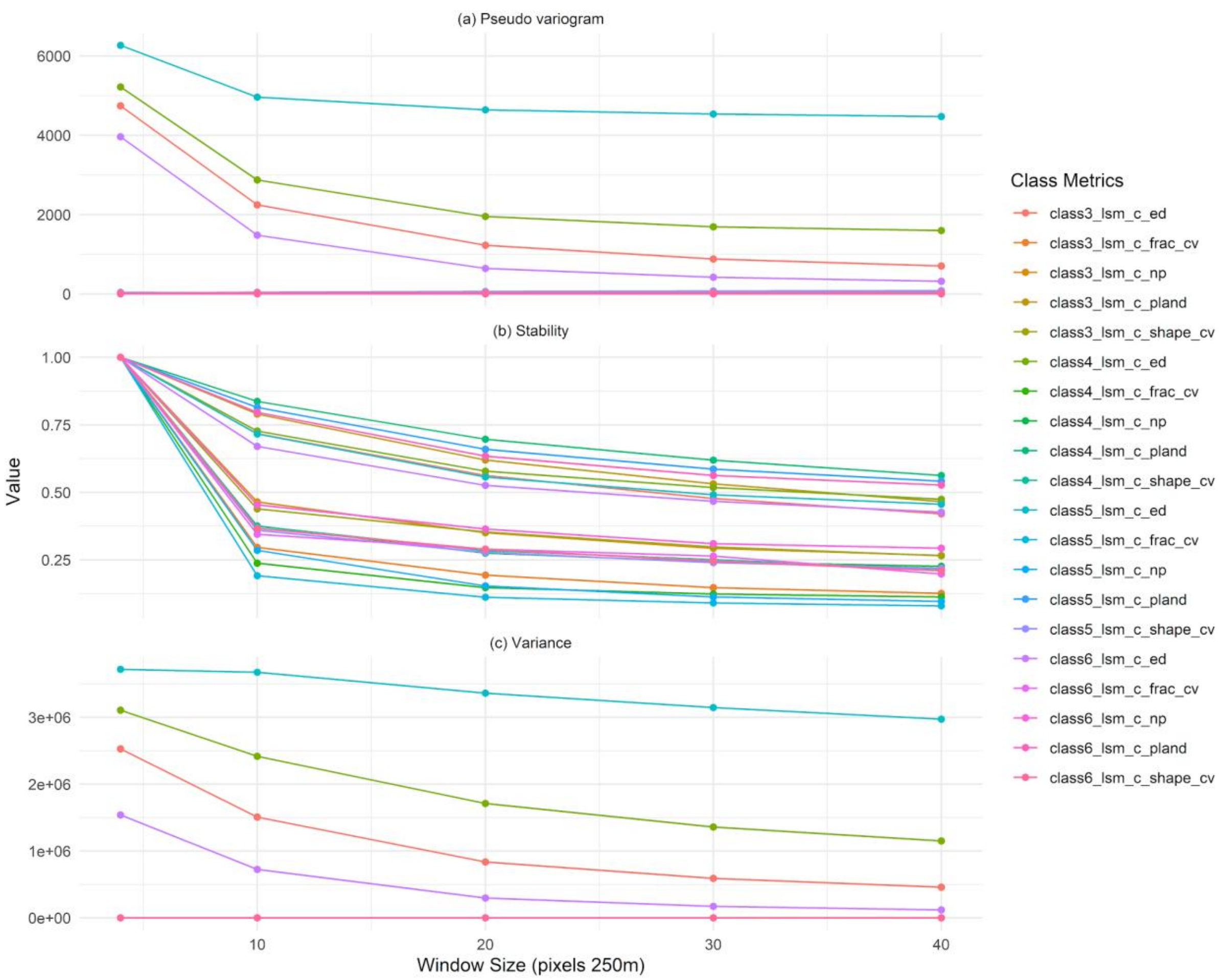

**Figure S3.** Selection of the optimal focal window size for calculating the spatial structure metrics of anthropogenic LULC classes (class 3 – Forest plantation, class 4 – Crop, class 5 – Pasture, class 6 – Urban area) was performed using three sensitivity tests according to window size (in pixels at a 250 m resolution): (a) Pseudo-variogram illustrating the mean spatial roughness of the metrics; (b) Inter-scale stability of the metrics assessed through correlation with the smallest window (4 pixels – 1 km); (c) Spatial variance of the metrics. The cross-scale analysis indicates that the spatial structure of the landscape changes markedly between 1 km (4 pixels) and 5 km (20 pixels).

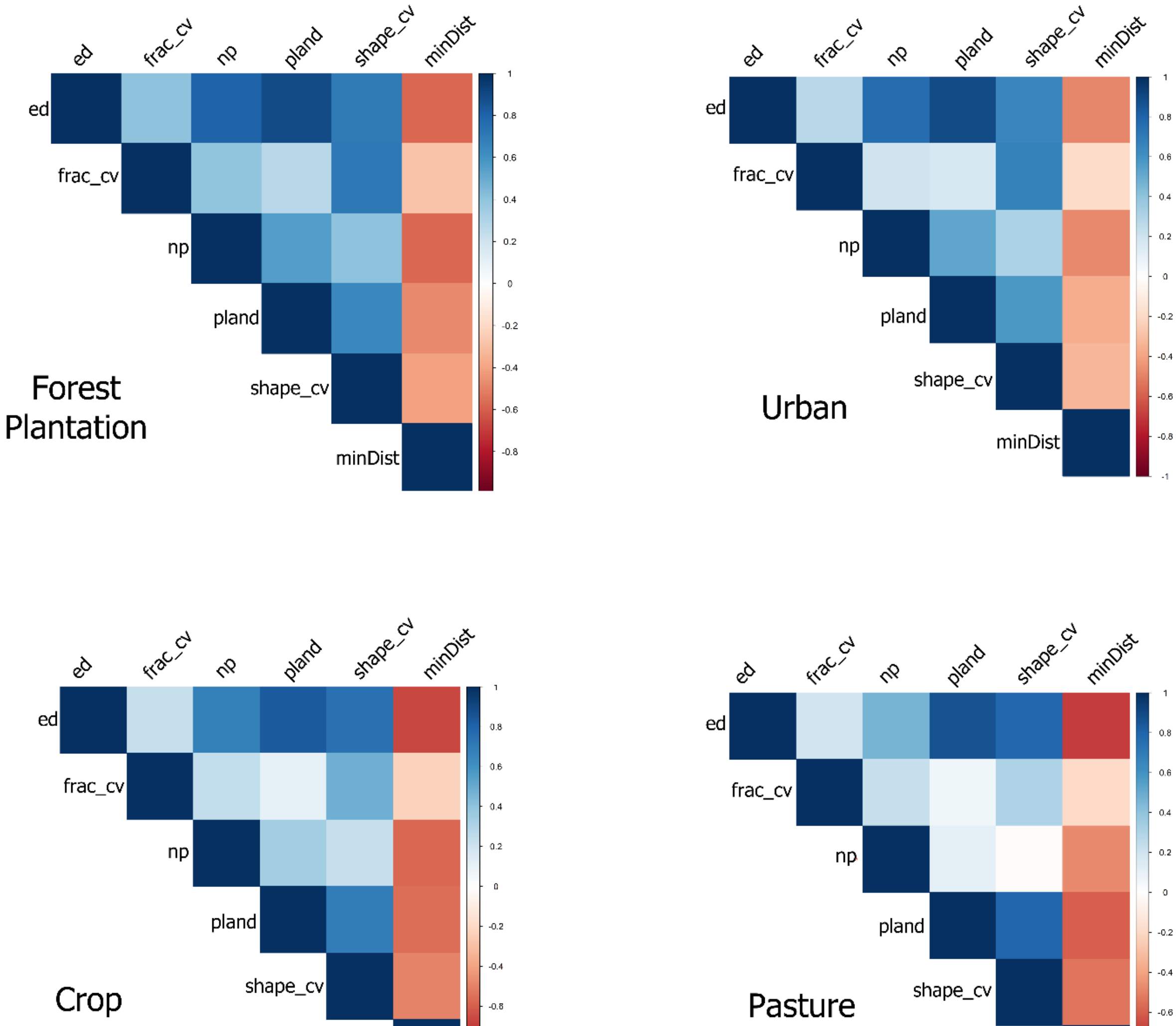

**Figure S4.** Corrplots showing pairwise correlations (Pearson's r) among class-level landscape metrics computed for four anthropogenic land-use/land-cover classes of interest (urban area, forest plantation, cropland, pasture). These correlation matrices were used to perform a collinearity analysis and to retain a subset of non-collinear and complementary metrics (selection criterion: $r<0.8$) that are widely recognised as informative in the landscape ecology literature, in order to robustly characterise the relative importance and influence of each anthropogenic class on landscape structure. The same metric-selection procedure was applied to all anthropogenic classes to ensure that the resulting class-specific pressure indices are comparable across classes and can subsequently be combined. Metrics displayed are: edge density (ED), number of patches (NP), percentage of landscape occupied by the class (PLAND), fractal dimension index (coefficient of variation; FRAC_CV), shape index (coefficient of variation; SHAPE_CV), and minimum distance to the class (minDist).

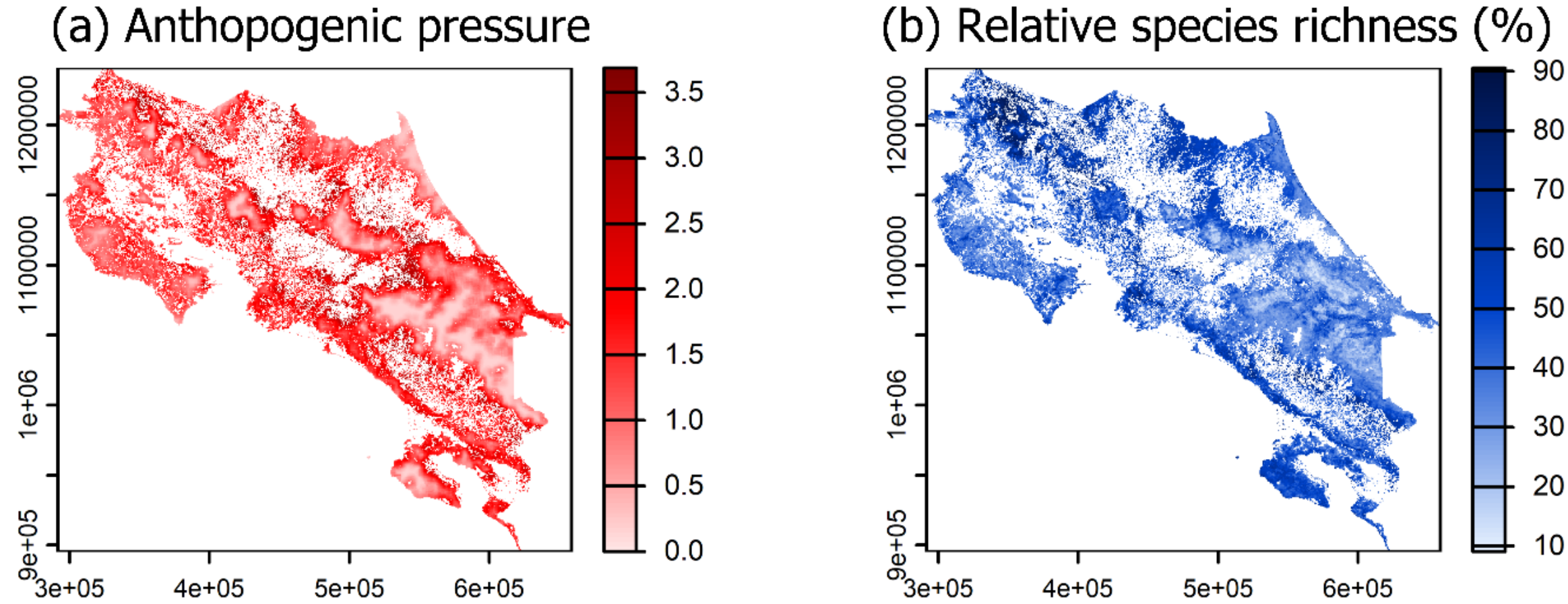

**Figure S5.** Mapping of (a) aggregated anthropogenic pressure at the landscape scale (within a 5 km radius around each forest pixel), obtained by combining the four pressure indices, and (b) Relative species richness, corresponding to percentage of studied species exhibiting high suitability scores (level 1 and 2 hotspots) within each forest pixel. The local anthropogenic pressure index (a) reveals a strong and continuous accumulation of pressures across a large part of the central and northern regions of the country, as well as in the northern Pacific area. In contrast, the mountainous regions of Talamanca, the forests of the Osa Peninsula, and some isolated Caribbean areas exhibit low to moderate pressure levels. The map of the percentage of species displaying high potential hotspots (b) highlights a clear concentration of hotspots in the mountain ranges, primarily the Guanacaste Range, as well as in the Osa Peninsula and the Fila Costeña Range. These regions show up to 90.55% of species in potential hotspots. Conversely, the Central Valley, the plains along the Pacific and Caribbean coasts, and certain areas of the Central and Talamanca mountain ranges exhibit very low levels of multi-species potential hotspots, with a minimum of 8.66% of species.

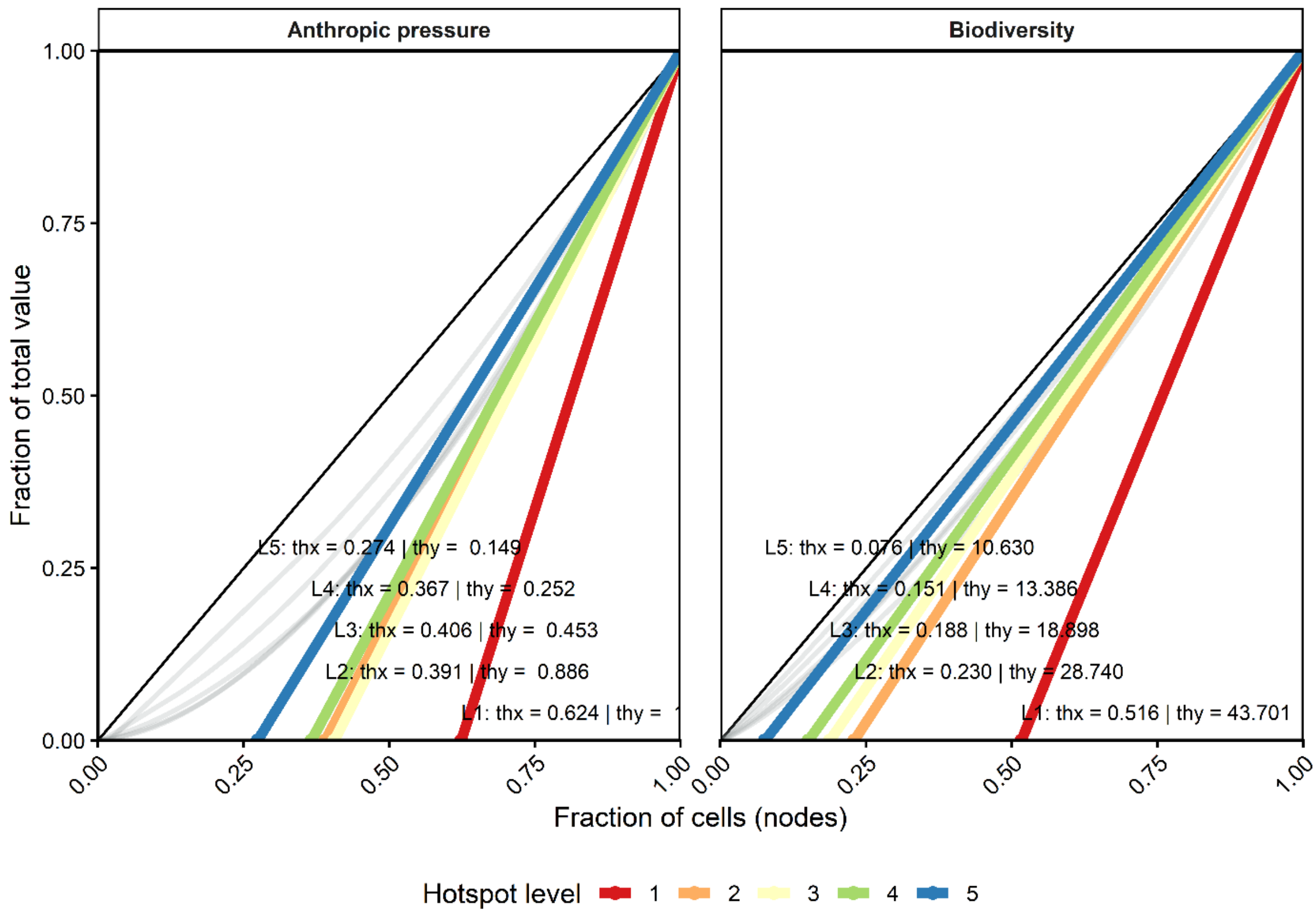

**Figure S6.** Lorenz curves used for Loubar discretisation of continuous relative species richness and anthropogenic pressure maps into six hotspot levels (level 1 = highest values; level 6 = lowest values).